\documentclass[aps,prd,superscriptaddress,nofootinbib,twocolumn,longbibliography]{revtex4-1}
\usepackage{amsmath,amssymb}
\usepackage{epsfig,graphicx}
\usepackage{xcolor}

\renewcommand{\theequation}{\arabic{section}.\arabic{equation}}
\renewcommand{\thesection}{\arabic{section}}

\catcode`@=11
\@addtoreset{equation}{section}
\catcode`@=12

\DeclareMathOperator{\sign}{sign}
\DeclareMathOperator{\arcoth}{arcoth}
\DeclareMathOperator{\arsinh}{arsinh}
\DeclareMathOperator{\tr}{tr}

\usepackage{hyperref}
\hypersetup{
    colorlinks,%
    citecolor=blue,%
    filecolor=blue,%
    linkcolor=blue,%
    urlcolor=purple
}

\begin{document}

\title{\bf Generalized unimodular gravity as a form of k-essence}
\author{A. O. Barvinsky}
\email{barvin@td.lpi.ru}
\affiliation{Theory Department, Lebedev
Physics Institute,
Leninsky Prospect 53, Moscow 119991, Russia}

\author{N. Kolganov}
\email{nikita.kolganov@phystech.edu}
\affiliation{Moscow Institute of Physics and Technology, Institutskii per. 9, Dolgoprudny 141700, Russia}
\affiliation{Institute for Theoretical and Mathematical Physics, Moscow State University, Leninskie Gory, GSP-1, 119991 Moscow, Russia}

\author{A. Vikman}
\email{vikman@fzu.cz}
\affiliation{CEICO, Institute of Physics of the Czech Academy of Sciences, Na Slovance 1999/2, 182 21 Prague 8, Czech Republic}


	\begin{abstract}
We consider modifications of general relativity characterized by a special  noncovariant constraint on metric coefficients, which effectively generates a perfect-fluid type of matter stress tensor in Einstein equations. Such class of modified gravity models includes recently suggested generalized unimodular gravity (GUMG) theory and its simplest version -- unimodular gravity (UMG). We make these gravity models covariant by introducing four Stueckelberg fields and show that in the case of generalized unimodular gravity three out of these fields dynamically decouple. This means that the covariant form of generalized unimodular gravity is dynamically equivalent to k-essence theory with a specific Lagrangian which can be reconstructed from the parameters of GUMG theory. We provide the examples, where such reconstruction can be done explicitly, and briefly discuss theories beyond GUMG, related to self-gravitating media models. Also we compare GUMG k-inflation with cuscuton models of dynamically inert k-essence field and discuss motivation for GUMG coming from effective field theory.
	\end{abstract}

\maketitle

\section{Introduction}
	
Dark energy and dark matter phenomena in cosmology and fundamental problems of quantum gravity form a rich playground for modifications of Einstein general relativity (GR). Such modifications are almost always associated with adding new physical degrees of freedom. Even when it is not done explicitly, one can often reformulate the resulting theory as GR interacting with additional degrees of freedom. For instance, this is the case for $R^2$-inflation \cite{Starobinsky} and other $f(R)$ theories, see \cite{Whitt:1984pd} and \cite{Schmidt:2001ac} accordingly. 

Another way to modify GR is to (partially) break general covariance like it is done in some formulations of the unimodular gravity \cite{vanderBij:1981ym,Buchmuller:1988wx,UMG_HT,UMG_Unruh}, Lorentz symmetry violating Horava gravity \cite{Horava}, massive gravity theories \cite{massive_fp,Rubakov:2008nh,deRham:2014zqa},  ``minimally modified gravity theories''  \cite{Lin:2017oow}, scalar-tensor theories in unitary gauge \cite{Gao:2014soa} and in a wide class of aether \cite{aether} or the so-called self-gravitating media models \cite{self-gravitating}, just to mention a few.
	
Breaking of general covariance can be enforced by including explicitly non-covariant terms to the action, for example, terms with purely algebraic dependence on metric coefficients. Such terms can be either explicitly added to the action or as an algebraic constraint on metric coefficients with the Lagrangian multiplier. In both cases the relevant part of the action can generate non-trivial stress-energy tensor. In the first case, usually associated with self-gravitating media models, this results from metric variation of manifestly noncovariant terms, whereas in the second case these are the Lagrange multiplier terms which enforce the constraints by the Lagrange multipliers procedure. This latter case and its peculiarities compared to self-gravitating media theories will be a major focus of this paper.
	
From the point of view of cosmological applications and effective hydrodynamical description of condensed media models, the most interesting type of the stress tensor is the perfect fluid one (which generically arises in the lowest order of gradient expansion)
	 \begin{equation}
	 T_{\mu\nu} = p \, g_{\mu\nu} + (p + \rho) u_\mu u_\nu.
	 \end{equation}
It is composed of density $\rho$ and pressure $p$ scalar functions and the local rest frame 4-velocity $u^{\mu}$, normalized to unity in spacetime metric, $g^{\mu\nu}u_\mu u_\nu=-1$. This local rest frame provides a natural (3+1)-dimensional spacetime split $\mu=(0,i)$, where $i=1,2,3$ denote spatial components orthogonal to $u^{\mu}$. As shown in \cite{self-gravitating}, metric  terms in the action, that generate such stress tensors of isotropic  $O(3)$-invariant perfect fluids, can depend only on two possible pairs of combinations of metric coefficients: $(g^{00}$, $\det g_{\mu\nu})$ and $(\det g^{ij}$, $\det g_{\mu\nu})$, where $g^{ij}$ are the spatial components of the {\em contravariant} spacetime metric. Correspondingly for metric constrained models this means that they are compatible with isotropic $O(3)$-invariant perfect fluids for the following two choices of metric constraints
	 \begin{equation}
	 V(g^{00}, (-g)^{-1/2}) = 0  \label{V_constr}
	 \end{equation}
and
	 \begin {equation}
	 V((\det g^{ij})^{1/2}, (-g)^{-1/2}) = 0,  \label{W_constr}
    \end{equation}
where $V$ is some local function of its two field arguments and $g=\det g_{\mu\nu}$. Qualitatively the difference between these two sets of constraints can be seen in terms of ADM variables \cite{Arnowitt:1962hi,Poisson} associated with the (3+1)-foliation of spacetime by spacelike hypersurfaces having the 4-velocity $u_\mu$ as their unit normal vector. The constraint (\ref{V_constr}) implies that the ADM lapse function ${\cal N}=(-g^{00})^{-1/2}$ expresses as a function of the determinant of the 3-dimensional metric, ${\cal N}=N(\gamma)$, $\gamma\equiv\det\gamma_{ij}$, $\gamma_{ij}\equiv g_{ij}$. On the contrary, the second constraint (\ref{W_constr}) is more complicated since it gives additional dependence\footnote{It should be emphasized that $g^{ij}$ in (\ref{W_constr}) is not the inverse of $\gamma_{ij}$, but rather depends on shift vectors, $g^{ij}=\gamma^{ij}-{\cal N}^i{\cal N}^j/{\cal N}^2$, ${\cal N}^i\equiv\gamma^{ij}{\cal N}_j$, $\gamma_{ij}\gamma^{jk}=\delta^k_i$.} of the lapse ${\cal N}=N(\gamma_{ij},{\cal N}_i)$ on shift functions ${\cal N}_i=g_{0i}$. 	
General relativity, equipped by the first type of constraint is nothing but recently proposed generalized unimodular gravity theory (GUMG) \cite{GUMG}. This generalization of a well-known unimodular gravity (UMG) is rather interesting on general theoretical grounds as an example of Hamiltonian systems with bifurcating constraints \cite{GUMG-can,HT-book} and, at the same time, an an example of the so-called unfree gauge systems \cite{Lyakhovich, Lyakhovich_BV}. Moreover, GUMG theory has successful applications within cosmological acceleration and inflation phenomena \cite{GUMG-can,GUMG-infl} and possible associations with the renormalizable projectible version of Horava gravity \cite{Horava,Horava-ren}. In particular, additional degree of freedom induced by diffeomorphism invariance violation originates in GUMG from purely metric sector and can be considered as a scalar graviton mode which in cosmology can play the role of inflaton.
	
As in other models with spontaneously broken symmetries, it is always possible to embed GUMG theory into a fully covariant framework by means of additional fields. For instance, UMG theory can be covariantized by means of a vector field \cite{UMG_HT,Jirousek:2018ago}, via four Stueckelberg fields \cite{UMG_Kuchar} or in terms of a gauge field \cite{UMG_Vikman}. Examples of covariantization also include Lorentz-invariant formulation of Horava gravity by means of the chronon field \cite{chronon} and covariant versions of massive gravity in terms of Stueckelberg scalar fields \cite{massive_dubovsky,Rubakov:2008nh,deRham:2014zqa}. Covariantization can obviously be done for any theory by introducing the same number of coordinate Stueckelberg fields as the spacetime dimensionality, but really interesting is the situation when the number of these auxiliary fields is not excessive and coincides with the number of extra physical degrees of freedom. Indeed, Horava gravity has only one additional mode and can be covariantized using a single chronon field, which restores coordinate-dependent temporal diffeomorphism invariance \cite{chronon}. purpleAt the same time, Stueckelberg covariantization of massive or self-gravitating media models does not reveal the additional dynamical degrees of freedom, as their number in such theories is less than the number of the Stueckelberg fields. Nevertheless, a covariantization is always useful, as it allows to use arbitrary physically well motivated coordinates associated with external test bodies and not with some degrees of freedom intrinsic to the system.

The GUMG action has a structure similar to those of self-gravitating media models and can be covariantized via four Stueckelberg scalar fields $\phi^A$, $A=(0,a)$, $a=1,2,3$.
Nevertheless, as we show in this paper, it turns out that three out of them, corresponding to spatial coordinates, completely decouple in the equations of motion from the temporal Stueckelberg field $\phi=\phi^0$ and become irrelevant. The resulting model can be reformulated as a well-known generally covariant k-essence theory \cite{k-infl} of this scalar field $\phi$ with the factorized Lagrangian of the form
	\begin{equation}
	\mathcal L = K(\phi)\,P(X),\quad\quad
	X\equiv g^{\mu\nu}\partial_\mu\phi\,\partial_\mu\phi,   \label{X}
	\end{equation}
	where the both functions $K(\phi)$ and $P(X)$ can be reconstructed from the constraint (\ref{V_constr}). This field is a generally covariant implementation of the non-covariant scalar graviton mode in the original formulation of GUMG theory. Remarkably, this formulation elucidates a typical k-essence expression for the speed of sound of the $\phi$ field perturbation modes on Friedmann background, which was found in \cite{GUMG-can,GUMG-infl} but could not be explained within a non-covariant formulation of GUMG.
	
This work is organized as follows. In Sect.2 we recapitulate the details of GUMG theory and its inflation theory application. Sect.3 contains the procedure of GUMG covariantization in terms of four coordinate Stueckelberg fields. In Sect.4 we show the decoupling of spatial Stueckelberg fields from the gravitational dynamics of spacetime metric and temporal Stueckelberg component $\phi^0$ playing from now on the role of k-essence. The k-essence Lagrangian is reconstructed from the constraint function of the original GUMG theory in Sect.5 on the background of Friedmann metric. This section also contains examples of this reconstruction including the case of GUMG inflation, which is treated within gradient and field expansion. The final section is devoted to conclusions and discussion. Appendix contains the case of multi-field k-essence reconstruction for a special class of constrained gravity theories motivated by models of self-gravitating media.

\section{Generalized unimodular gravity} \label{sec_GUMG}
Generalized unimodular gravity theory \cite{GUMG,GUMG-can} differs from the original unimodular gravity by replacing the condition $\sqrt{-g} = 1$ with a more general condition on metric coefficients
	\begin{equation}
		(-g^{00})^{-1/2} = N(\gamma), \label{GUMG_constr}
	\end{equation}
where $N(\gamma)$ is a rather generic function of the determinant of the spatial metric $\gamma = \det g_{ij}$. In contrast to UMG, this constraint breaks not only diffeomorphism invariance but also Lorentz invariance. If one enforces this constraint by adding it to the Einstein-Hilbert action $S_\text{EH}[g_{\mu\nu}]$ action with the Lagrange multiplier $\Lambda$
	\begin{equation}
	S_\text{GUMG}[g_{\mu\nu}, \Lambda] =
    S_\text{EH}[g_{\mu\nu}]+\! \int d^4x \, \Lambda \Bigl(\frac1{\sqrt{-g^{00}}}
    - N(\gamma)\Bigr), \label{action_GUMG_noncov}
	\end{equation}
then this action generates as equations of motion the Einstein equations with the matter stress tensor of the effective perfect fluid
	\begin{equation}
		T_{\mu\nu} = (\rho + p) \, g_{\mu\nu} + p \, v_\mu v_\nu,
	\end{equation}
which has the following parameters of density, pressure, barotropic equation of state and 4-velocity \cite{GUMG,GUMG-can}
	\begin{equation}
	\rho=-\frac\Lambda{2\sqrt\gamma},\,\,
	p = w \rho, \,\,
	w = 2\frac{d \ln  N(\gamma)}{d \ln \gamma}, \,\,
	v_\mu = \frac{\delta^0_\mu}{\sqrt{-g^{00}}}. \label{pf_pars}
	\end{equation}
It is important that with the Lagrange multiplier obtained from tracing the equations of motion, these ten variational equations become linearly dependent projections of the {\em vacuum} Einstein equations. This means that every solution of vacuum GR is also a solution of GUMG \cite{GUMG-can}, but of course not vice versa. GUMG theory has the additional degree of freedom -- the scalar graviton which can generate the analogue of the inflationary cosmology with phenomenologically acceptable parameters \cite{GUMG-infl}.

The simplest manifestation of this degree of freedom is the presence of dark fluid density $\rho$ in the Friedmann equation for the scale factor $a(t)$ in the FRW metric $ds^2 = -{\cal N}^2 dt^2+a^2(t)\delta_{ij}dx^i dx^j$,
    \begin{eqnarray}
    H^2=\frac{\rho}{3M_P^2},\quad
    \rho=\frac{C}{{\cal N}a^3}.            \label{C}
    \end{eqnarray}
Here $H=\dot a/{\cal N}a$ is the Hubble factor and $C$ is the constant of integration of the spatial $ij$-components of Einstein equations corresponding to the homogeneous mode of this scalar (conformal) graviton degree of freedom.\footnote{We consider the case of spatially flat Friedmann metric with $k=0$ and use the units with reduced Planck mass $M_P=1/\sqrt{8\pi G}$.} As shown in \cite{GUMG-can,GUMG-infl}, the dark fluid term of (\ref{C}) can be interpreted as dark energy in cosmological acceleration scenario with the red shift dependent barotropic parameter $w$ (which was one of the motivations for GUMG theory) and, what is perhaps more promising, it can be interpreted as a source of inflation and inflationary CMB spectra generated entirely by the metric sector of the theory. Moreover, the theory of GUMG cosmological perturbations has many features of the hydrodynamical description formalism of inflation developed in \cite{Mukhanov_book} employing k-essence theory \cite{k-infl,k-pert}, characterized by a nontrivial speed of sound of hydrodynamical media excitations.

In particular, the canonically normalized physical mode of the scalar graviton $\vartheta$ in the typical decomposition of cosmological perturbations into irreducible components, $\delta\gamma_{ij}= a^2(-2 \psi\,\delta_{ij}+ 2 \partial_i \partial_j E+2\partial_{(i} F_{j)}+t_{ij})$, has on the above Friedmann background the quadratic action \cite{GUMG-infl}
    \begin{eqnarray}
    &&S = \frac12 \int d\eta \,
    d^3x \Bigl( \vartheta'^2 + c_s^2 \,
    \vartheta \Delta \vartheta
    + \frac{\theta^{\prime\prime}\!\!}{\theta}
    \,  \vartheta^2 \Bigr),                   \label{2.6}\\
    &&\vartheta = \theta \, \psi, \qquad \theta^2
    = 3 a^2 M_P^2 \frac{\Omega}{w},         \label{theta_psi}\\
    &&c_s^2 = \frac{w(1+w)}{\Omega},
    \qquad \Omega=1+ w+2 \frac{d\ln w}{d\ln\gamma}  \label{gumg_pars_pert}
    \end{eqnarray}
(it is written in the conformal time parametrization, $\eta=\int dt\,{\cal N}/a$, prime denoting the $\eta$-derivative $d/d\eta$). This action generates for $\vartheta$ the well-known Mukhanov-Sasaki equation \cite{k-pert,Mukhanov_book} with a nontrivial speed of sound $c_s$ defined by the background quantities $w$ and $\Omega$. In the domain of parameters free of ghost and gradient instabilities,
    \begin{equation}
    \frac{w}\Omega>0, \quad 1+w>0,     \label{stability_domain}
    \end{equation}
$c_s^2>0$ and this equation has oscillatory modes. For a given comoving scale $k=|{\mathbf k}|$ in the regime of quasi-exponential cosmological expansion, $w\simeq -1$, these modes $\vartheta_{\mathbf k}(\eta)$ give rise to the primordial spectrum of scalar perturbations
$\delta_\psi^2(k,\eta) = k^3|\vartheta_{\mathbf k}(\eta)|^2/2\pi^2\theta^2(\eta)$, which gets formed at the point of horizon crossing, $c_s k = H a$. As shown in \cite{GUMG-infl}, this spectrum is given by the same expression as in the hydrodynamical formalism of the inflation theory \cite{Mukhanov_book},
    \begin{equation}
    \delta_\psi^2(k, \eta) = \frac1{36\pi^2}\frac1{c_s(1+w)}
    \frac\rho{M_P^4}\biggr|_{c_s k = H a},
    \end{equation}
in terms of the parameters of the effective perfect fluid -- $\rho$, $w$ and $c_s$ --- taken at the same horizon crossing point. Under the following choice of the function $N(\gamma)$ in the metric constraint (\ref{GUMG_constr})
	\begin{equation}
	N(\gamma) = \frac1{\sqrt{\gamma}}
    \Bigl[1 + \Bigl(\frac{\gamma}{\gamma_*}\Bigr)^{\frac12}+ B \, \Bigl(\frac{\gamma}{\gamma_*}\Bigr)^{\frac32
    + \frac43(n_s-1)}\Bigr],                  \label{N_of_gamma_CMB}
	\end{equation}
scalar power spectrum fits well the observable CMBR data with a small red tilt $n_s-1\simeq -0.04$ and small tensor to scalar ratio $r=(e^N)^{-4(1-n_s)}\sim 0.001$ which is expressible in GUMG model in terms of the usually accepted e-folding number $N\simeq 60$ \cite{GUMG-infl}. In (\ref{N_of_gamma_CMB}) $B$ is some positive $O(1)$ parameter and $\gamma_*$ corresponds to the value of the 3-metric determinant at the end of inflation stage, so that the e-folding number $N=N(k)$ of the horizon crossing point for a scale $k$ is defined by the relation $e^N=(\gamma_*/\gamma)^{1/6}|_{c_s k = H a}$.\footnote{In view of this the ansatz (\ref{N_of_gamma_CMB}) should be understood as the expansion in positive powers of the ratio $\gamma/\gamma_*<1$, in which only the first few terms have been retained. It makes sense at the inflation stage $\gamma\ll\gamma_*$ and, of course, cannot be extrapolated beyond the end of inflation point.}

Despite these strong parallels between GUMG theory and hydrodynamical formulation of inflation and k-inflation, their congruence was not found in \cite{GUMG-can,GUMG-infl} to be complete. In particular, the speed of sound naively defined from the equation of state as $c_s^2=dp/d\rho$ does not reproduce the correct answer (\ref{gumg_pars_pert}). Correct expression (\ref{gumg_pars_pert}) should be mediated in relativistic formulation of perfect fluid hydrodynamics by the dependence of both $p$ and $\rho$ on the relativistic invariant $X=g^{\mu\nu}\partial_\mu\phi\,\partial_\mu\phi$ of some k-essence field $\phi$, so that the correct speed of sound should read $c_s^2=\partial_X p/\partial_X\rho$. For this purpose this k-essence field should be constructed within a covariant version of GUMG theory, which is the subject of the next section.
	
	\section{Stueckelberg method of GUMG theory covariantization} \label{sec_cov}

Consider the theory of gravity, which differs from General Relativity by some algebraic constraint imposed on the metric coefficients and written in the form $U(g^{\mu\nu}) = 0$. This constraint can be implemented in the action of the theory $S$ by the Lagrange multipliers procedure
	\begin{equation}
	S[g_{\mu\nu}, \Lambda] = S_\text{EH}[g_{\mu\nu}]
    + \int d^4x\sqrt{-g} \, \Lambda \, U(g^{\mu\nu}). \label{action_mod_noncov}
	\end{equation}
GUMG model belongs to this class of modified gravity theories, where the constraint has a special form (\ref{GUMG_constr}).
	
The action (\ref{action_mod_noncov}) does not have the property of general covariance for a generic function $U(g^{\mu\nu})$, which is obvious for all ultralocal functions $U(g^{\mu\nu})$ because there are no algebraic combinations of metric coefficients that transform as scalars. To restore general covariance one can use a well-known Stueckelberg method. For this purpose we introduce four Stueckelberg fields $\phi^A(x)$, $A = 0, \ldots 3$, which mimic the change of coordinates $x^\mu \mapsto \phi^A(x)$. Obviously, Einstein-Hilbert term and measure $d^4x \sqrt{-g}$ are invariant under such a replacement, while the metric argument of the function $U(g^{\mu\nu})$ goes over into
	\begin{equation}
	g^{\mu\nu} \mapsto C^{AB}, \qquad C^{AB}
    = g^{\mu\nu} \partial_\mu \phi^A \partial_\nu \phi^B, \label{cov_rule}
	\end{equation}
and the action (\ref{action_mod_noncov}) takes the form
	\begin{equation}
	S[g_{\mu\nu}, \Lambda, \phi^A] = S_\text{EH}[g_{\mu\nu}]
    + \int d^4x \sqrt{-g} \, \Lambda \, U(C^{AB}). \label{action_mod_cov}
	\end{equation}
Now, if one treats Stueckelberg fields $\phi^A$ and the Lagrange multiplier $\Lambda$ as scalars, the resulting action becomes generally covariant under diffeomorphisms of $g_{\mu\nu}(x)$, $\phi^A(x)$ and $\Lambda(x)$. In the gauge $\phi^A(x) = x^A$ the action $S[g_{\mu\nu}, \Lambda, \phi^A]$ takes the original noncovariant form (\ref{action_mod_noncov}).
	
The equivalence of the actions (\ref{action_mod_noncov}) and (\ref{action_mod_cov}) follows from the fact that the variation of Stueckelberg fields does not lead to new dynamics, i.e. equations of motion $\delta S/ \delta \phi^A = 0$ follow from the variation with respect to the metric coefficients and the Lagrange multiplier. To demonstrate this property, write down the full set of equations of motion, namely Einstein equations with the following stress-energy tensor
	\begin{equation}
	T_{\mu\nu} = \Lambda \, U \, g_{\mu\nu}
    - 2 \, \Lambda \frac{\partial U}{\partial C^{AB}}
    \partial_\mu \phi^A \partial_\nu \phi^B,    \label{set_cov}
	\end{equation}
covariant form of the constraint equation
	\begin{equation}
		U(C^{AB}) = 0, \label{constr_cov}
	\end{equation}
and the equations of motion for Stueckelberg fields $\phi^A$
	\begin{equation}
	\nabla^\mu \left(\Lambda
    \frac{\partial U}{\partial C^{AB}}
    \nabla_\mu \phi^B \right) = 0. \label{stuck_eom}
	\end{equation}
Bianchi identities lead to the conservation of the stress-energy tensor (\ref{set_cov}) $\nabla^\mu T_{\mu\nu} = 0$. Together with the constraint equation (\ref{constr_cov}) it gives
	\begin{equation}
	\nabla^\mu T_{\mu\nu} = -2 \nabla_\nu \phi^B \,
    \nabla^\mu \left(\Lambda \frac{\partial U}{\partial C^{AB}}
    \nabla_\mu \phi^A \right) = 0, \label{conservation}
	\end{equation}
so that if $\det \nabla_\nu \phi^B \ne 0$, the Stueckelberg fields equations (\ref{stuck_eom}) follow from Einstein equations and the constraint equation.

The model with the action (\ref{action_mod_cov}) is very similar to the theory of self-gravitating media (at leading order of gradient expansion) with the action of the form
	\begin{equation}
    S_\text{SGM}[g_{\mu\nu}, \phi^A] = S_\text{EH}[g_{\mu\nu}] + \int d^4x \sqrt{-g} \, U(C^{AB}),
    \end{equation}
where $C^{AB}$ is defined as in (\ref{cov_rule}). The main difference between self-gravitating media theory and the {\em constrained} gravity (\ref{action_mod_cov}) is the presence of additional Lagrange multiplier field $\Lambda$ in the second term (additional to the Einstein-Hilbert term) of the latter. Therefore, the matter stress-energy tensors of these theories differ by the overall factor $\Lambda$, that is $T_{\mu\nu} = \Lambda (T_\text{SGM})_{\mu\nu}$, where $(T_\text{SGM})_{\mu\nu}$ is a stress tensor of self gravitating media.

This difference does not change the classification of Lagrangians by the symmetry principle, adopted for self-gravitating media models \cite{self-gravitating}. As it was mentioned in Introduction, there are only two classes of functions of $C^{AB}$ that lead to $SO(3)$-invariant theory whose stress-energy tensor coincides with the stress tensor of some perfect fluid. Theories that have this property are most interesting from the viewpoint of cosmological applications. Namely, the function $U(C^{AB})$ in such theories has the following two types of $C^{AB}$-dependence
	\begin{equation}
	U(C^{AB}) = V(X, Z) \label{V_constr_cov}
	\end{equation}
and
	\begin{equation}
	U(C^{AB}) = V(b, Z), \label{W_constr_cov}
	\end{equation}
where we use the following conventional notation of self-gravitating media models \cite{self-gravitating}
	\begin{equation}
	X = C^{00}, \quad Z = \sqrt{-\det C^{AB}},
    \quad b = \sqrt{\det C^{ab}},
	\end{equation}
with the indices $a,b$ labelling the ``spatial'' components of the Stueckelberg fields
	\begin{equation}
	\phi^A=(\phi^0,\phi^a), \quad a = 1,2,3.
	\end{equation}

Obviously, multiplication of the stress-energy tensor by an overall scalar multiplier $\Lambda$ neither changes its perfect fluid form nor breaks its $SO(3)$ symmetry. Therefore, the constrained theories (\ref{action_mod_cov}) necessarily have the function $U(C^{AB})$ of the form (\ref{V_constr_cov}) or (\ref{W_constr_cov}), if one requires its stress-energy tensor to be of a $SO(3)$-invariant perfect fluid form, just as in the self-gravitating media case. Fixing the gauge $\phi^A(x) = x^A$, corresponding to the transition to a non-covariant version of the theory (\ref{action_mod_noncov}), immediately brings the functions $V(X, Z)$ and $V(b, Z)$ respectively to the forms (\ref{V_constr}) and (\ref{W_constr}), since in this gauge
	\begin{equation}
	X = g^{00},\,\,Z = (-g)^{-1/2},\,\,b = (\det g^{ij})^{1/2}.
	\end{equation}

\section{Decoupling of spatial Stueckelberg fields and k-essence}
Interesting feature of the constrained gravity theories (\ref{action_mod_cov}) and their non-covariant version (\ref{action_mod_noncov}), belonging to the family of functions (\ref{V_constr_cov}) and (\ref{W_constr_cov}), is a dynamical decoupling of a certain part of their Stueckelberg fields. Here we demonstrate this property for the case of (\ref{V_constr_cov}) corresponding to GUMG models. For the case of (\ref{W_constr_cov}) we refer the reader to Appendix \ref{app}.
	
The constraint (\ref{V_constr}) and it's covariant version $V(X, Z) = 0$ can be implemented in various equivalent ways. The most appropriate choice is based on the assumption that the constraint can be resolved with respect to the variable $Z$ as $Z = P(X)$, so that we set $V(X, Z) = P(X) - Z$, and the action becomes
	\begin{equation}
	S[g_{\mu\nu}, \Lambda, \phi^A]
    = S_\text{EH}[g_{\mu\nu}]
    + \int d^4x \sqrt{-g} \,
    \Lambda \, (P(X) - Z). \label{action_GUMG_cov}
	\end{equation}
In order to relate the function $P(X)$ to the function $N(\gamma)$ in the non-covariant version of the constraint (\ref{GUMG_constr}) one should covariantize its constituents $g^{00}$ and $\gamma$ by using the rule (\ref{cov_rule})
	\begin{equation}
	g^{00} \mapsto X, \quad \gamma=g \, g^{00} \mapsto -X / Z^2. \label{cov_rule_ADM}
	\end{equation}
Then the covariant form of the GUMG constraint (\ref{GUMG_constr}) takes the form $(-X)^{-1/2} = N(-X/Z^2)$, and its solution with respect to $Z$ reads
	\begin{equation}
	Z=P(X),\quad P(X) \equiv \left(\frac{-X}{\Gamma(1/\sqrt{-X})}
    \right)^{1/2},        \label{GUMG_constr_cov}
    \end{equation}
where $\Gamma(N)$ is the function inverse to $N(\gamma)$,
	\begin{equation}
	\Gamma(N(\gamma))\equiv\gamma.
    \end{equation}

The stress tensor corresponding to the action (\ref{action_GUMG_cov}) reads
	\begin{equation}
	T_{\mu\nu} = \Lambda \,(P g_{\mu\nu} + 2 P_X X v_\mu v_\nu),
    \quad v_\mu =
    -\frac{\partial_\mu \phi^0}{\sqrt{-X}}, \label{set_GUMG_cov}
	\end{equation}
where the subscript $X$ denotes a partial derivative with respect to $X$, $P_X=\partial_X P(X)$. As expected, it has the form of a perfect fluid stress tensor with the hydrodynamical parameters
	\begin{equation}
	p = w \rho =\Lambda \,P,\quad w = \frac{P}{2 P_X X - P}. \label{perf_fl}
	\end{equation}

Remarkably, this metric stress tensor, which serves as a matter source in the right hand side of Einstein equations, does not at all involve spatial components of Stueckelberg fields $\phi^a$, $a = 1,2,3$. As it was shown above, Stueckelberg field equations do not lead to additional dynamics and follow from Einstein equations. The constraint equation (\ref{GUMG_constr_cov}) only constrains spatial Stueckelberg fields $\phi^a$ by the temporal one $\phi^0$. This means that the only dynamical equations are Einstein equations, which do not depend on spatial components $\phi^a$. In this sense Stueckelberg fields $\phi^a$ decouple from the dynamics of the gravitational sector of the theory.
	
One might think that they actually couple via additional variable which enters the expression (\ref{set_GUMG_cov}) for the stress tensor -- the Lagrange multiplier $\Lambda$. However, the following reasoning shows that this variable can be regarded as a function of only the temporal component $\phi^0$. Indeed, use Bianchi identities to derive the conservation law of the stress tensor (\ref{set_GUMG_cov}) $\nabla^\mu T_{\mu\nu} = 0$, project it onto the linear space orthogonal to the perfect fluid 4-velocity $v_\mu$, and obtain
	\begin{equation}
	(\delta^\nu_\sigma + v^\nu v_\sigma) \nabla^\mu T_{\mu\nu}
    = P(X) \,(\delta^\nu_\sigma + v^\nu v_\sigma)
    \partial_\nu \Lambda = 0, \label{cons_proj}
	\end{equation}
whence $\partial_\nu \Lambda \propto \partial_\nu \phi^0$.\footnote{Alternatively, this directly follows from the variational equations of motion (\ref{stuck_eom}) for spatial Stueckelberg fields $\phi^a$ with the function $U(C^{AB})=P(X)-Z$.} Then it uniquely follows that $\Lambda = K(\phi^0)$ is a function of one variable $\phi^0$ -- this can be easily seen by the regular change of coordinates from $x^\mu$ to $\phi^A(x)$ under which $\Lambda(x)=K(\phi^A(x))$ and $\partial K(\phi^A)/\partial\phi^a=0$.\footnote{Thus {\em on-shell} value of $\Lambda$ can depend on spacetime coordinates only via $\phi^0(x)$. Note, however, that generally $\Lambda$ can still depend on those combinations of $\phi^a$ which are independent of $x$, including in particular spatially constant constants of motion, which means that $K(\phi^A(x))$ can depend on initial conditions, which will be confirmed by concrete examples considered in the next section.}

Moreover, it is easy to see that under this identification, $\Lambda(x)=K(\phi^0(x))$, the left hand side of the equation of motion (\ref{stuck_eom}) for $\phi^0$ with the function $U(C^{AB})=P(X)-Z$
	\begin{equation}
	\frac1{\sqrt{-g}}\frac{\delta S}{\delta \phi^0}
    = -2\nabla_\mu\bigl(\Lambda P_X\nabla^\mu\phi^0\bigr)
    +Z\,
    \partial_\mu\Lambda\,\bigl(\partial\phi^{-1}\bigr)^\mu_0
	\end{equation}
(here $(\partial\phi^{-1})^\mu_0$ are the elements of the matrix inverse to $\partial_\mu\phi^A$) can be rewritten as
	\begin{equation}
	\frac1{\sqrt{-g}}\frac{\delta S}{\delta \phi^0}
    = \left(\frac{\partial}{\partial \phi^0}
    - \nabla_\mu \frac{\partial}{\partial (\nabla_\mu\phi^0)}\right) K(\phi^0) P(X). \label{k-ess_eom}
	\end{equation}
This relation holds only in virtue of the constraint $Z=P(X)$ and the corollary of the spatial components, $A=a$, of the equations of motion (\ref{stuck_eom}) -- it does not rely on the equation of motion $\delta S/\delta\phi^0=0$ for $\phi^0$ itself.

Therefore, the equations of motion of the theory (\ref{action_GUMG_cov}) are equivalent to the equations of motion of the scalar field $\phi=\phi^0$ minimally coupled to General Relativity, whose action is
	\begin{equation}
	S_K[g_{\mu\nu}, \phi] = S_\text{EH}[g_{\mu\nu}]
    + \int d^4x \sqrt{-g} \, K(\phi) P(X). \label{action_k-ess}
	\end{equation}
Indeed, the variation of this action with respect to $\phi$ leads to (\ref{k-ess_eom}), while variation of the same action with respect to the metric field leads to Einstein equations with the stress-energy tensor
	\begin{equation}
	T_{\mu\nu} = K(\phi^0) P(X)\, g_{\mu\nu}
    + 2 K(\phi^0) P_X(X)\, X v_\mu v_\nu, \label{k-ess_set}
	\end{equation}
which is the result of substitution of $\Lambda = K(\phi^0)$ into the stress tensor (\ref{set_GUMG_cov}).

The theory with the action (\ref{action_k-ess}) has the form of a well-known k-essence theory. It is the statement of the dynamical equivalence between GUMG and k-essence theory -- one can forget about spatial Stueckelberg fields and drop a zero label of their temporal component which is identified with k-essence field $\phi$.

This explains the observation made in the papers \cite{GUMG-can,GUMG-infl},
where the GUMG perturbation theory on Friedmannian background was first developed, that this perturbation theory strongly resembles cosmological perturbation theory of a well-known k-essence theory. As we see now, the equivalence between (\ref{action_GUMG_cov}) and (\ref{action_k-ess}) extends beyond perturbation theory. However, this equivalence relies on a concrete form of the function $K(\phi)$, which arises only on shell -- as the result of solving all equations of motion for the theory with the covariantized GUMG action (\ref{action_GUMG_cov}). Unfortuately, the k-essence action cannot be regarded as the result of integrating out spatial Stueckelberg fields $\phi^a$ and the Lagrange multiplier $\Lambda$ in (\ref{action_GUMG_cov}), because this multiplier enters the action linearly and the relevant equations for these fields cannot be resolved for them. The nature of equivalence between (\ref{action_GUMG_cov}) and (\ref{action_k-ess}) is more complicated, because determination of $\Lambda$ as a function $K(\phi)$ requires to use Einstein equations from the metric sector. This, in particular, results in a special property of the k-essence Lagrangian -- its function $K(\phi)$ starts depending on initial conditions of the theory, which enter this function in the process of its reconstruction from equations of motion. In Sect.\ref{sec_k-ess} below we demonstrate this in the cosmological setup on Friedmann background.
	
\subsection{Conserved charge}
Covariant formulation in terms of Stueckelberg fields has the translation invariance property under constant shifts of $\phi^A$, $\phi^A(x)\mapsto\phi^A(x)+c^A$, which leads to local conserved currents $J^\mu_A$ --- their conservation being just the equations of motion (\ref{stuck_eom}),
	\begin{equation}
	J^\mu_A = -2\Lambda \frac{\partial U}{\partial C^{AB}}
    \nabla^\mu \phi^B,\quad \nabla_\mu J^\mu_A = 0.  \label{current}
	\end{equation}
In view of the constraint equation $U=0$ these currents are trivially related to the stress tensor $T_{\mu\nu}=\partial_\mu\phi^A J_{A\nu}$, so that $J^\mu_A=(\partial\phi)^{-1\nu}_{\;\;\;\;A}\,T^\mu_\nu$ \footnote{Despite this representation the current conservation is associated not with the usual Killing nature of this vector, but follows from the different mechanism --- the constraint on metric coefficients $U=0$. Indeed, differentiation of (\ref{current}) gives $\nabla^\mu J_{\mu A}=\Lambda (\partial\phi)^{-1\mu}_{\;\;\;\;A}\,\partial_\mu U=0$ --- the alternative form of Eq.~(\ref{conservation}).}. They obviously generate conserved charges $Q_A$ which can be obtained by integrating the current continuity equation over a spacetime domain. Consider the barrel type domain $\cal M$ of points $x$, bounded by the top and bottom ``spherical'' patches of spacelike hypersurfaces --- 3-dimensional balls of radius $R$,
	\begin{equation}
	\sigma_\pm:\;\;\;\phi(x)=t_\pm,\;\; \phi^a(x)\delta_{ab}\phi^b(x)
    \leq R^2,
	\end{equation}
and the segment of the cylindrical timelike hypersurface --- the side boundary of $\cal M$,
	\begin{equation}
	\sigma_\vdash:\;\;\;t_-\leq\phi(x)\leq t_+,\;\;
    \phi^a(x)\delta_{ab}\phi^b(x)=R^2.
	\end{equation}
Integration gives
	\begin{equation}
	0=\int\limits_{\cal M}d^4x\,\sqrt{-g}\,\nabla_\mu J^\mu_A
    =Q_A^+-Q_A^-+F_A,     \label{bulk_int}
	\end{equation}
where $Q_A^\pm$ are the charges associated with the top and bottom boundaries of $\cal M$ and $F_A$ are the current fluxes through the side boundary $\sigma_\vdash$,
	\begin{eqnarray}
	Q_A^\pm=\int\limits_{\sigma_\pm} d^3\sigma_\mu\,J^\mu_A,\quad
    F_A=\int\limits_{\sigma_\vdash} d^3\sigma_\mu\,J^\mu_A.
	\end{eqnarray}
Here $d^3\sigma_\mu$ are the future pointing integration elements of the surfaces $\sigma_\pm$ and the outward pointing element of the side boundary $\sigma_\vdash$. Obviously, the charge conservation should follow from the vanishing of the flux $F$, but in contrast to usual systems, in which one freely assumes vanishing fields at spatial infinity, here the situation is nontrivial, because the Stueckelberg fields linearly grow under a natural assumption of asymptotical spatial flatness and flat spacetime foliation, $\phi^A(x)\to x^\mu\times\delta^A_\mu$.

Nevertheless, in the case of the action (\ref{action_GUMG_cov}) with the constraint function $U=P(X)-\sqrt{-\det C^{AB}}$ the situation remains manageable. Moreover, as we will see, in this case the set of four charges $Q_A$ degenerates to just one charge $Q_0$, and this holds in full accordance with the decoupling of three spatial Stueckelberg fields. Indeed, using this particular expression for $U$ one finds on shell
	\begin{eqnarray}
	Q_A=\int\limits_{\sigma} d^3x \sqrt\gamma\,v_\mu J^\mu_A=\delta^0_A
    \int\limits_{\sigma} d^3x \sqrt\gamma\,\mathcal N\rho,
	\end{eqnarray}
where we used the chain of relations
	\begin{eqnarray}
	v_\mu J^\mu_A=\delta^0_A \Lambda\,\frac{2XP_X-P}{\sqrt{-X}}=
    \delta^0_A\, \mathcal N\rho,
	\end{eqnarray}
based on the constraint $P(X)=\sqrt{-\det C^{AB}}$, Eq.(\ref{perf_fl}) and the expression for the lapse function. Therefore, only the temporal charge $Q_0$ survives, whereas the spatial charges $Q_a$ identically vanish. 

The situation with corresponding fluxes $F_A$ is different --- $F_0$ is vanishing for $R\to\infty$ (provided $\phi^A(x)$ sufficiently quickly tending to $x^\mu\delta^A_\mu$), while $F_a$ are a priori nonzero. This can be seen by converting the bulk integration in (\ref{bulk_int}) to new coordinates -- foliation time $t$, $t=\phi(x)$, and three ``cartesian'' spatial coordinates $\boldsymbol\phi=\phi^a$ coinciding with spatial Stueckelberg fields $\phi^a=\phi^a(x)$. By reparameterizing the latter to spherical coordinates -- the radius $r=(\boldsymbol\phi^2)^{1/2}$ and the angles of the unit vector $n^a=\phi^a/r$ on the 2-dimensional sphere, one has the relation between the bulk integration measures
	\begin{equation}
	\sqrt{-g}\,d^4x=\frac{dt\,d^3\boldsymbol\phi}{\sqrt{-\det C^{AB}}}
    =\frac{r^2}{P}\,dr\,dt\,d^2\Omega,
	\end{equation}
where $d^2\Omega$ is invariant integration measure on a 2-sphere. It follows then that
	\begin{equation}
	F_A=R^2\int\limits_{t_-}^{t_+}dt
    \left.\int d^2\Omega\,\frac{J^r_A}{P}\right|_{r=R}, \label{flux}
	\end{equation}
where 
	\begin{equation}
    J^r_A=\partial_\mu r \, J^\mu_A=-2\Lambda P_X \, \delta^0_A C^{0a}n_a+
    \Lambda P \, \delta^a_A n_a   \label{Jrad}
	\end{equation}
is the radial component of the vector current. For $A=0$ this expression reduces to the first term which is proportional to $\partial_\mu\phi^0g^{\mu\nu}\partial_\nu\phi^a\to 0$, $R\to\infty$, for asymptotically flat systems with $\phi^A(x)$ sufficiently quickly tending to $x^\mu\delta^A_\mu$ at infinity. This implies vanishing $F_0$ in (\ref{bulk_int}) and the conservation of the temporal charge,
	\begin{eqnarray}
	Q_0=\int\limits_{R\to\infty} d^3x \sqrt\gamma\,\mathcal N\rho,
	\end{eqnarray}
the total energy of the perfect fluid\footnote{For rotating systems with slowly decreasing shift functions $C^{0a}$ this flux is nonzero and implies via Eq.(\ref{bulk_int}) the rate of the energy change by accretion.  Note also that the integrand of $Q_0$ is not the geometrically invariant energy density $\sqrt\gamma\rho$, but rather $\mathcal N\sqrt\gamma\rho$, the factor $N$ indicating the gravitational defect of the total mass of matter \cite{Landafshitz_Teorpol}.}.
For $A=a$ the flux (\ref{flux}) seems a priory being nonzero, which contradicts identically vanishing charges $Q_a^\pm$ in (\ref{bulk_int}). However, on substitution of (\ref{Jrad}) into (\ref{flux}) one has
	\begin{equation}
	F_a=R^2\int\limits_{t_-}^{t_+}dt\,\Lambda(t)
    \int d^2\Omega\,n_a=0,
	\end{equation}
where in the coordinate system of $(t,\phi^a)$ a spatially homogeneous $\Lambda(t)=K(\phi)|_{\phi=t}$ leads to the vanishing flux as a result of angular integration. This confirms complete consistency of our Stueckelberg decoupling procedure and the k-essence dependence of the Lagrange multiplier $\Lambda=K(\phi)$.

\section{Reconstruction of k-essence} \label{sec_k-ess}
The explicit form of the function $K(\phi)$ should be found from the full set of equations of motion for the action (\ref{action_GUMG_cov}), which is technically possible in quadratures on the Friedmann background
	\begin{equation}
	\begin{aligned}
	&ds^2 = - \mathcal N^2(t)\, dt^2 + a^2(t)\, \delta_{ij} dx^i dx^j,\\
	&\phi = \phi(t), \quad \phi^a = x^a,
    \quad \Lambda = \Lambda(t). \label{bg_fr}
	\end{aligned}
	\end{equation}
Here $\phi^a=x^a=x^i$ is the only choice compatible with spatial homogeneity of the perfect fluid stress tensor on the homogeneous Friedmann background. On this background the action (\ref{action_GUMG_cov}) becomes
	\begin{align}
	S[a, \mathcal N, \phi, \Lambda]=&
	-3 M_P^2 \int dt \, \mathcal N a^3
    \frac{\dot a^2}{\mathcal N^2 a^2}\nonumber\\
    &+ \int dt \, \mathcal N \, \Lambda\, \bigl(a^3 P(X)
    - \sigma \sqrt{-X}\bigr), \\
	\sigma\equiv &\sign(\dot \phi / \mathcal N),
    \qquad X = -\frac{\dot{\phi}^2}{\mathcal N^2}. \nonumber
	\end{align}
We note that the sign factor $\sigma$ remains constant during a continuous evolution of $\phi$, because changing the sign of $\dot \phi / \mathcal N$ condradicts the nonvanishing value of $\det \partial_\mu \phi^A =\dot \phi$, $\phi=\phi^0$, which we assume in the equivalence of GUMG and k-essence theories.

The constraint equation corresponding to the variation of $\Lambda$ gives the scale factor $a$ as a function of $X$
	\begin{equation}
	a^3 = \sigma \, \sqrt{-X}/P(X), \label{constr_fr}
	\end{equation}
	which inversely can be used to express $X$ as the function of $a$, $X=X(a)$.

The variational equation of motion for $\phi$ gives
	\begin{equation}
	-\Lambda \, (2 \sigma a^3 P_X \sqrt{-X} + 1) = C,   \label{Fried_const}
	\end{equation}
where $C$ is the integration constant determined by initial conditions. One can express $\Lambda$ from the last equality and use the definition (\ref{perf_fl}) of $w$ along with the constraint equation (\ref{constr_fr}) to find that
	\begin{equation}
		\Lambda = C \, w(X). \label{Lambda_w}
	\end{equation}
Using this relation, one can rewrite Friedmann equation
	\begin{equation}
	\frac{\dot a^2}{\mathcal N^2 a^2} =
    \frac{\Lambda}{3 M_P^2} (2P_X X - P)
	\end{equation}
as
	\begin{equation}
	\frac{d \ln a^3}{\mathcal N dt} = M_P^{-1}\sqrt{3 C P(X)}.
	\end{equation}
Dividing it by $\dot \phi/\mathcal N = \sigma \sqrt{-X}$ and substituting $X(a)$ -- the solution of the constraint equation (\ref{constr_fr}) for $X$, one finds a differential equation on $\phi(a)$, which can be directly integrated to give
	\begin{align}
	\phi(a) - \phi_0 &= M_P\int_{a_0}^{a} da'\,a'^2 \,
    \sqrt{\frac{3P(X(a'))}C} \nonumber\\
	&\equiv\Phi(a) - \Phi(a_0).        \label{phi_of_a}
	\end{align}
Without loss of generality one can choose $\phi_0 = \Phi(a_0)$ and express $a=a(\phi)$ as a function of $\phi$, so that it can be further used in $X(a)$. Therefore, the relation (\ref{Lambda_w}) finally yields $\Lambda$ as a function of $\phi$
	\begin{equation}
	\Lambda = C w(X(a(\phi))) \equiv K(\phi). \label{K_def}
	\end{equation}
This equality determines the desired function $K(\phi)$ which is the part of the k-essence action (\ref{action_k-ess}).

As in the case of a general background, when solving the k-essence equations of motion, one should use the same initial conditions as those used in the derivation of the function $K(\phi)$. Namely, these equations comprise a second order differential equation on the scalar field $\phi$, which requires initial values of the field and its time derivative, and a first order differential equation on the scale factor, which requires the initial value of the scale factor. Initial values of the scalar field $\phi_0$ and the scale factor $a_0$ should be related by (\ref{phi_of_a}), while the initial velocity of the scalar field implicitly follows from the constraint (\ref{constr_fr}). Therefore, given $a(t_0) = a_0$, the remaining inital conditions read
	\begin{equation}
	\phi(t_0) = \Phi(a_0), \qquad
    \frac{\dot \phi}{\mathcal N}\biggr|_{t_0}
    = \sigma \sqrt{-X(a_0)}                    \label{k-ess_in_cond}
	\end{equation}

	We show now that perfect fluid parameters and inflationary perturbations in the covariantized (\ref{action_GUMG_cov}) and k-essence (\ref{action_k-ess}) theories are equivalent to those of Sect.\ref{sec_GUMG}, obtained in \cite{GUMG,GUMG-can,GUMG-infl} within the original non-covariant formalism. First of all, on the constraint surface (\ref{GUMG_constr_cov}) covariantization rule (\ref{cov_rule_ADM}) for $\gamma$ takes the form of $\gamma= \Gamma(1/\sqrt{-X})$.	Therefore, the barotropic parameter $w$ defined by Eq.(\ref{perf_fl}) can be written as
	\begin{align}
		w &=\left[\frac{d}{d\ln X}\ln\frac{P^2}X\right]^{-1} = 2\left[\frac{d\ln\Gamma(1/\sqrt{-X})}{d\ln(1/\sqrt{-X})}\right]^{-1}  \nonumber\\
    	&= \left.2\,\frac{d \ln N(\gamma)}{d \ln \gamma}\right|_{\gamma = \Gamma(1/\sqrt{-X})},
	\end{align}
	which exactly coincides with (\ref{pf_pars}). The action of linearized scalar perturbations of k-essence has the same form as (\ref{2.6}) where the parameters $\theta$ and $c_s$ equal \cite{Mukhanov_book,k-pert}
	\begin{equation}
		c_s^2 = \frac{\partial p/\partial X}{\partial\rho/\partial X}, \quad
		\theta^2 = a^2 \frac{1+w}{c_s^2}.
	\end{equation}
	We should compare these parameters with (\ref{gumg_pars_pert}), obtained from non-covariant version of GUMG. Using hydrodynamical parameters (\ref{perf_fl}) and the definition (\ref{GUMG_constr_cov}) of $P(X(\gamma))=1/\sqrt\gamma N(\gamma)$ via $\gamma$ and $N(\gamma)$, we obtain
	\begin{eqnarray}
		c_s^2 &=& \frac{\partial p/\partial X}{\partial\rho/\partial X} =
    	\frac{\partial_\gamma P}{\partial_\gamma(P/w)}\biggr|_{\gamma= \Gamma(1/\sqrt{-X})}\nonumber\\
    &=& \frac{w (1 + w)}{\Omega}\biggr|_{\gamma =\Gamma(1/\sqrt{-X})},
	\end{eqnarray}
which coincides with the result (\ref{gumg_pars_pert}) of \cite{GUMG-can,GUMG-infl}.\footnote{Note that $\Lambda=K(\phi)$ in $p=\Lambda P$ and $\rho=\Lambda P/w$ is not differentiated with respect to $X$ and completely cancels out.} Similarly, for the canonical normalization parameter $\theta$ we have
	\begin{equation}
	\theta^2 = 3 M_P^2 \,a^2 \frac{1 + w}{c_s^2} = 3M_P^2 \, a^2
    \frac{\Omega}{w}\biggr|_{\gamma =\Gamma(1/\sqrt{-X})},
	\end{equation}
which also coincide with (\ref{gumg_pars_pert}). Equality of parameters of the linearized action implies that primordial power spectra also coincide in both versions of the theory. Now, we apply this procedure of $K$-reconstruction from a given $P(X)$ for several particular examples.
	
\subsection{Exactly solvable example}
Consider the function $P(X)$ of the form
	\begin{equation}
	P(X) = \frac{-\alpha X}{\beta^2 - \sqrt{-X}} \label{ex1_P_of_X}
	\end{equation}
with constant parameters $\alpha$ and $\beta$. This choice is interesting because all steps of the reconstruction procedure (along with integration of the k-essence equations of motion) can be explicitly implemented in terms of elementary functions. The constraint (\ref{constr_fr}) can be solved as
	\begin{equation}
		\sqrt{-X(a)} = \frac{\beta^2}{1 + \alpha \, a^3}, \label{ex1_X_of_a}
	\end{equation}
so that the integration in Eq.(\ref{phi_of_a}) gives
	\begin{equation}
	\phi(a) = \tilde \phi\,\arsinh \sqrt{\alpha a^3},
    \quad
    \tilde \phi \equiv \frac{2\beta M_P}{\sqrt{3C\alpha}},
	\end{equation}
whence
	\begin{equation}
	a^3(\phi) =\frac1{\alpha}
    \sinh^2\frac\phi{\tilde\phi}.\label{ex1_a_of_phi}
	\end{equation}
To use (\ref{K_def}), we first calculate the barotropic parameter (\ref{perf_fl}) for this $P(X)$ defined by (\ref{ex1_P_of_X}) via an obvious chain of relations
	\begin{equation}
	\begin{aligned}
		w &= \frac{P}{2 P_X X - P} =
    \frac1{\beta^2} (\beta^2 - \sqrt{-X}) \\
    &= \frac{\alpha \, a^3}{1 + \alpha \, a^3}
    = \tanh^2\frac\phi{\tilde\phi}        \label{w_forms}
	\end{aligned}
	\end{equation}
and finally obtain the following k-essence Lagrangian
	\begin{align}
	\mathcal L_K(\phi, X) &= K(\phi) P(X) \nonumber \\
    &= C  \tanh^2\frac\phi{\tilde\phi}\,
    \frac{-\alpha X}{\beta^2 - \sqrt{-X}}          \label{k-ess_part}
	\end{align}
Thus, as it has been mentioned above, the initial conditions explicitly enter the k-essence Lagrangian. According to (\ref{k-ess_in_cond}) equations generated by this Lagrangian should be used with these initial conditions parameterized by $a(t_0) = a_0$ and the constant $C$,
	\begin{equation}
	\phi(t_0) = \tilde \phi \, \arsinh \sqrt{\alpha a_0^3},
    \,\,\,\, \frac{\dot \phi}{\mathcal N}\biggr|_{t_0} = \frac{\beta^2}{1 + \alpha \, a_0^3}.
	\end{equation}

\subsection{Example of the canonically normalizable mode}

The second example corresponds to
	\begin{equation}
		P(X) = -\beta X - \alpha \label{ex2_P_of_X}
	\end{equation}
for constant parameters $\alpha$ and $\beta$. This example can be interesting because the field redefinition
	\begin{equation}
		\varphi = \int_0^\phi d\phi' \sqrt{2 \beta K(\phi')}
	\end{equation}
brings the scalar field part of the action (\ref{action_k-ess}) to the canonical normalization in the kinetic term of the Lagrangian
    \begin{equation}
	\mathcal L_\varphi =\frac12 \frac{\dot \varphi}{\mathcal N^2}
    -V(\varphi)
	\end{equation}
having the potential $V(\varphi) = - \alpha K(\phi(\varphi))$. This example looks closer to inflation theory whose slow roll regime seems to follow from steepness properties of this potential. This anticipation, however, turns out to be misleading because the initial conditions (\ref{k-ess_in_cond}) violate slow roll approximation -- as we will see, the kinetic term of the scalar field turns out to be large in the flatness domain of the potential $V(\varphi)$.

As for the field $\phi$, its canonically normalized counterpart $\varphi$ also has very specific initial conditions which immediately follow from the initial conditions (\ref{k-ess_in_cond}) on $\phi$ and the definition of $\varphi$, namely
 	\begin{align}
 	\varphi(t_0) &=
    \int_0^{\phi_0} d\phi' \sqrt{2 \beta K(\phi')} \label{ex2_in_cond_1} \\
 	\frac{\dot \varphi}{ \mathcal N}\biggr|_{t_0} &=
    \sqrt{2 \beta K(\phi_0)} \,
    \frac{\dot \phi}{ \mathcal N}\biggr|_{t_0}. \label{ex2_in_cond_2}
 	\end{align}
To keep kinetic and potential terms positive, one should choose $\alpha$ and $\beta$ of the same signs. The overall sign is irrelevant, so we choose positive $\alpha$ and $\beta$.	The constraint (\ref{constr_fr}) can be resolved as
	\begin{equation}
	\sqrt{-X_\pm(a)} = \frac1{2\beta \, a^3}(\sqrt{1
    + 4\alpha\beta \, a^{6}} \pm 1),    \label{ex2_X_of_a}
	\end{equation}
where the sign is selected, depending on the sign of $\alpha$ and $\beta$, from the requirement of positivity of $H^2\propto CP(X)$. The integral in Eq.(\ref{phi_of_a}) with the substitution of these $X_\pm(a)$ can be taken in elementary functions
	\begin{align}
	\Phi_+(a) &= \tilde\phi \, \bigl(2 u(a) -
    \arcoth u(a)\bigr),        \label{ex2_F_of_a_p} \\
    u(a) &= \frac1{\sqrt 2} \bigl(\sqrt{1 + 4 \alpha \beta \, a^6}+1\bigr)^{1/2}, \nonumber \\
	\Phi_-(a) &= \tilde\phi \,
    (2 v(a) - \arctan v(a)), \label{ex2_F_of_a_n} \\
    v(a) &= \frac1{\sqrt 2} \bigl(\sqrt{1 + 4 \alpha \beta \, a^6}-1\bigr)^{1/2},  \nonumber
	\end{align}
where $\tilde \phi \equiv M_P/\sqrt{3 |\beta C|}$ and the
subscripts correspond to the choice of the sign in (\ref{ex2_X_of_a}). However, the resulting dependence of $\phi(a)$ on $a$ cannot be inverted explicitly. For this reason we provide only the asymptotic behavior of $K(\phi)$ and the corresponding potential $V(\varphi)$.
	
\begin{figure}[h!]
	\centering
	\includegraphics[width=0.5\textwidth]{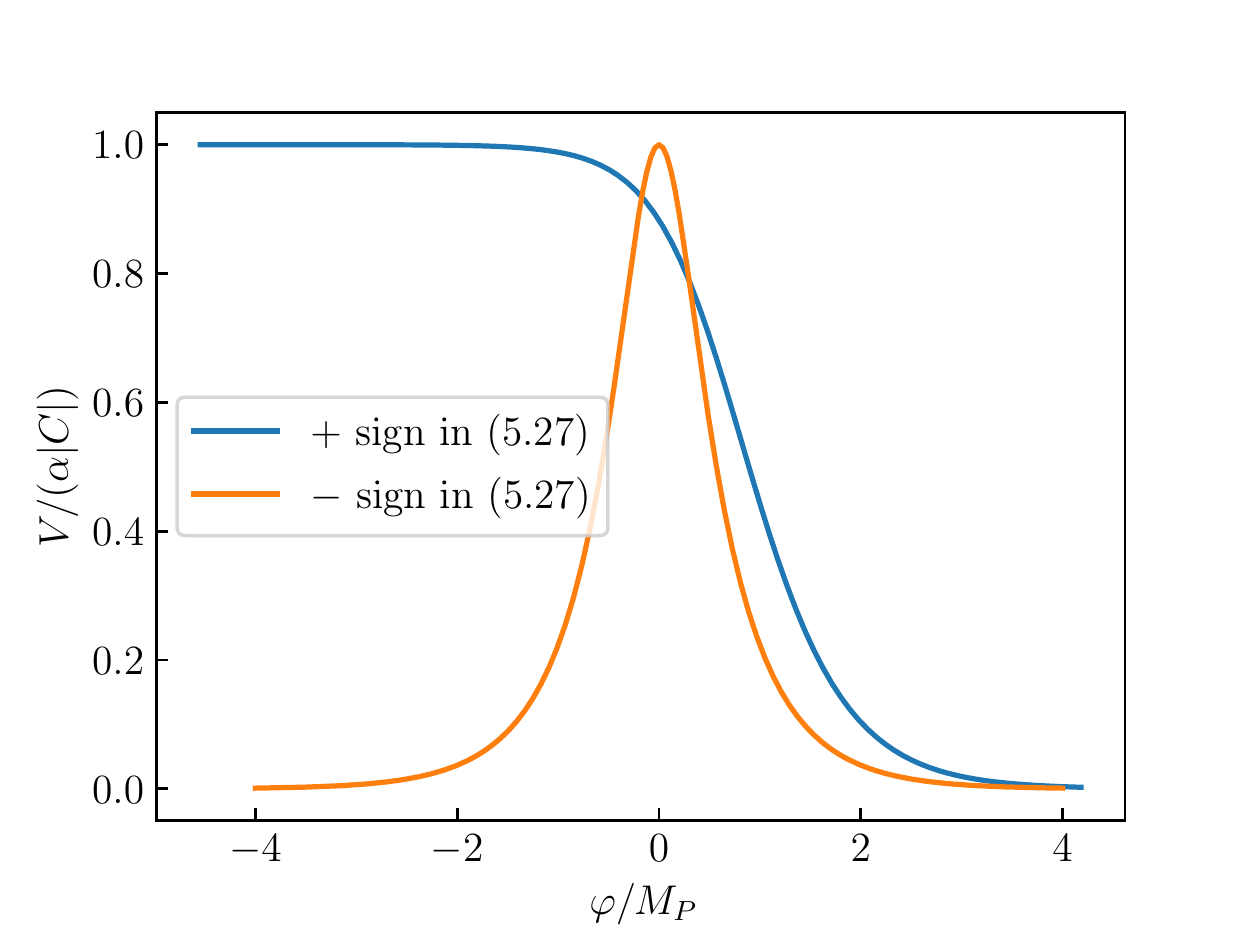}
	\caption{Modified gravity model (\ref{action_GUMG_cov}) with a
    particular $P(X)$ dependence (\ref{ex2_P_of_X}) on Friedmann background is equivalent to the minimally coupled scalar field with one of the depicted potentials and specific initial conditions (\ref{ex2_in_cond_1}) -- (\ref{ex2_in_cond_2}).}
		\label{fig_pot}
	\end{figure}

In the case of a positive sign in (\ref{ex2_X_of_a}) and $\phi / \tilde\phi \to -\infty $ the asymptotics read
	\begin{align}
	&K(\phi) = C + O\bigl[e^{-2|\phi| / \tilde\phi}\bigr],\\
	&\varphi = \sqrt{\frac23} \frac\phi{\tilde\phi} + O(1), \quad
    \frac\varphi{M_P} \to -\infty, \\
	&V(\varphi) = - C \alpha+ O\bigl[e^{-\sqrt{6}|\varphi|/M_P}\bigr], \label{ex2_pot_noninfl}
	\end{align}
whereas in the $\phi / \tilde\phi \to+\infty$ limit one has
	\begin{align}
	&K(\phi) =2C \Bigl(\frac\phi{\tilde\phi}\Bigr)^{-2}
    + O\Bigl[\Bigl(\frac\phi{\tilde\phi}\Bigr)^{-3}\Bigr], \\
	&\frac\varphi{M_P} = \frac2{\sqrt{3}} \ln\frac\phi{\tilde\phi} + O(1),
    \quad \frac\varphi{M_P} \to+\infty, \\
	&V(\varphi) = - C \alpha\, O\bigl[e^{-\sqrt{3}\varphi/M_P}\bigr].
	\end{align}
See blue line on Fig.\ref{fig_pot} for numerically calculated plot of the potential.
	
In the case of a positive sign in (\ref{ex2_X_of_a}) and $|\phi|/\tilde\phi \to 0$ the asymptotic behavior of the main quantities looks as follows
	\begin{align}
	&K(\phi) = C \Bigl(1 - 2\Bigl(\frac{\phi}{\tilde\phi}\Bigr)^2\Bigr)
    + O\Bigl[\Bigl(\frac{\phi}{\tilde\phi}\Bigr)^4\Bigr] \\
	&\frac{\varphi}{M_P} = \sqrt{\frac23} \frac{\phi}{\tilde\phi}
    + O\Bigl[\Bigl(\frac{\phi}{\tilde\phi}\Bigr)^3\Bigr], \hspace{0.5em} \frac{|\varphi|}{M_P} \to 0, \\
	&V(\varphi) = - C \alpha \Bigl(1 - 3 \Bigl(\frac{\varphi}{M_P}\Bigr)^{\!2}\Bigr)
    + O\Bigl[\Bigl(\frac{\varphi}{M_P}\Bigr)^{\!4}\Bigr]
	\end{align}
while in $|\phi|/\tilde\phi \to \infty$ limit one has
	\begin{align}
		&K(\phi) = -2C  \Bigl(\frac{\phi}{\tilde\phi}\Bigr)^{-2} + O\Bigl[\Bigl(\frac{\phi}{\tilde\phi}\Bigr)^{-3}\Bigr] \\
		&\frac{\varphi}{M_P} = \sign\phi \; \frac2{\sqrt{3}} \ln\frac{|\phi|}{\tilde\phi} + O(1), \quad \frac{|\varphi|}{M_P} \to \infty, \\
		&V(\varphi) = C \alpha \, O\bigl[e^{-\sqrt{3}|\varphi|/M_P}\bigr].
	\end{align}
See orange line on Fig. \ref{fig_pot} for numerically calculated plot of the potential.

From the blue line of the plot on Fig.\ref{fig_pot} one could have expected that this case provides slow roll inflation for large negative $\varphi$, $|\varphi| / M_P \to  \infty$, where the potential is flat and slow-roll parameters are small. Unfortunately, it is not the case, because according to Eq.(\ref{ex2_X_of_a}) one cannot choose initial kinetic energy arbitrary small unless the parameters are artificially tuned. Indeed, according to (\ref{ex2_F_of_a_p}), the scale factor is exponentially small in this region, $a_0^3 \simeq e^{-\sqrt{3/2} \, |\varphi| / M_P}$, so that the kinetic energy defined by (\ref{ex2_in_cond_2}) and (\ref{ex2_X_of_a}), $\dot\varphi^2 /(2\mathcal N^2) \simeq (C/\beta) \, a_0^{-6} \simeq (C/\beta) \, e^{\sqrt{6} |\varphi| / M_P}$, becomes at large $|\varphi|/M_P$ much larger than the approximately constant potential energy (\ref{ex2_pot_noninfl}).
	
\subsection{GUMG k-inflation}

Interesting example, which indeed allows for inflation theory applications, corresponds to the case of generalized unimodular gravity (\ref{action_GUMG_noncov}) with $N(\gamma)$ defined by (\ref{N_of_gamma_CMB}). In particular, it leads to cosmological perturbation theory which reproduces the primordial power spectrum consistent with basic CMB data \cite{GUMG-infl}. As in the previous example, this case is not exactly solvable, but within expansion in integer and fractional powers of the square root of the ratio $\gamma/\gamma_*$, which is small during a major part of inflationary stage, the dual k-essence formulation can be easily obtained.

To bring the theory to its covariant form (\ref{action_GUMG_cov}) we first solve (\ref{N_of_gamma_CMB}) with respect to $\gamma$ to find the function $\Gamma({\cal N})$ (inverse to ${\cal N}=N(\gamma)$) as an expansion in (fractional) powers of $1/{\cal N}\sqrt{\gamma_*}$, which after the substitution ${\cal N}=1/\sqrt{-X}$ leads to the expansion of $P(X)$ in powers of the new small parameter
	\begin{equation}
	\delta\equiv\Bigl(\frac{-X}{\gamma_*}\Bigr)^{\frac12}.
	\end{equation}
For the $P(X)$ given by Eq.(\ref{GUMG_constr_cov}) the first few terms of the resulting expansion read as
	\begin{equation}
	P(X) = 1 - \Bigl(\frac{-X}{\gamma_*}\Bigr)^{\frac12}
    - B \Bigl(\frac{-X}{\gamma_*}\Bigr)^{\frac32
    + 4\frac{n_s-1}3} + \ldots\,\,,                      \label{P_of_X_CMB}
	\end{equation}
where ellipses denote higher order powers  of $\delta$ and $\delta^{3+8\frac{n_s-1}3}$.

The further step -- the procedure of reconstructing $K(\phi)$ -- leads to another expansion in integer and fractional powers of the new smallness parameter
	\begin{equation}
	\varepsilon\equiv\frac{3H_0\phi}{\sqrt{\gamma_*}}.
	\end{equation}
Here $H_0$ is the initial value of the Hubble factor, which parameterizes the constant $C$ according to the Friedmann equation (\ref{C}), $M_P^2H_0^2\simeq C/3$.  At the beginning of inflation stage the scale factor $a_0$ is much smaller than its value at the end of inflation, $a_0/a_*=(\gamma_0/\gamma_*)^{1/6}< e^{-60}$, so that in view of the constraint (\ref{N_of_gamma_CMB}) ${\cal N}\sqrt\gamma\simeq 1$, which explains this parameterization.

Using the constraint (\ref{constr_fr}) and Eqs.(\ref{Fried_const}), (\ref{phi_of_a}) and (\ref{K_def}) one finds by iteration procedure that in the lowest order of this expansion $\sqrt{-X(a)}\simeq a^3$ and $a^3(\phi)\simeq 3H_0\phi$, while the expansion for $K(\phi)$ (with slightly higher accuracy) begins with
	\begin{align}
	K(\phi)= & -3M_P^2H_0^2 \Bigl[1-\varepsilon+\frac34\varepsilon^2
    -\frac12\varepsilon^3\nonumber\\
    &-\bigl(3+8\,\tfrac{n_s-1}3\bigr) B\,\varepsilon^{3+8\frac{n_s-1}3}+ \ldots \Bigr],
	\end{align}
where ellipses denote higher order powers  of $\varepsilon$ and $\varepsilon^{3+8\frac{n_s-1}3}$.

Altogether, the Lagrangian of the k-inflation model dual to our inflationary GUMG theory takes the form of both gradient and field expansion
	\begin{align}
	{\cal L}_K(&\phi,X)=\nonumber\\
    &\;3 M_P^2 H_0^2\,
    \sqrt{\frac{-X}{\gamma_*}}\,\Bigl[1+O\Bigl(\varepsilon,\varepsilon^{3+ 8\frac{n_s-1}3},\delta,\delta^{3+8\frac{n_s-1}3}\Bigr)\Bigr]\nonumber\\
    &-3 M_P^2 H_0^2\,\Bigl[1+O\Bigl(\varepsilon,\varepsilon^{3+ 8\frac{n_s-1}3}\Bigr)\Bigr].  \label{grad_exp}
	\end{align}
Equations of motion for the k-essence field with this Lagrangian should be used with the following initial conditions at the onset of inflation
	\begin{equation}
	\phi(t_0)\simeq \frac{a_0^3}{H_0}, \qquad
    \frac{\dot\phi}{\mathcal N}\biggr|_{t_0}
    =\sqrt{-X}\biggr|_{t_0} \simeq a_0^3.
	\end{equation}

Interesting feature of the expansion (\ref{grad_exp}) is that modulo $\delta$ and $\varepsilon$ corrections this is the Lagrangian of the so-called cuscuton model of \cite{cuscuton} with the nonlinear square-root kinetic term $\mu^4\sqrt{-X}$, $\mu^4\equiv 3 M_P^2 H_0^2/\sqrt{\gamma_*}$, and the constant potential $V\equiv 3M_P^2H_0^2$. This model is interesting because with a special choice of the potential cuscuton models are claimed in \cite{cuscuton} to generate dark energy mechanism similar to that of the Dvali-Gabadadze-Porrati model \cite{DGP}. Generically, cuscuton models are known for a peculiar property of their scalar field -- it is not dynamical, because its equation is ultralocal in time, which is interpreted as infinite speed of sound of the ``propagating'' excitations of $\phi$. Indeed, the cuscuton can be obtained as a limit of the superluminal (anti)DBI introduced in \cite{Mukhanov:2005bu,Babichev:2006vx,Babichev:2007wg}. This property is in full accordance with the fact that in GUMG model with the function (\ref{N_of_gamma_CMB}) truncated to its first two terms ($B=0$) the parameter $\Omega$ in (\ref{gumg_pars_pert}) vanishes, and the speed of sound $c_s$ diverges.

The third term of (\ref{N_of_gamma_CMB}) (and correspondingly in (\ref{P_of_X_CMB})) is critically important in GUMG model to make its scalar mode dynamical, render its speed of sound finite and be capable of generating small red tilt $n_s-1\simeq -0.04$ in the primordial CMB spectrum. Smallness of this tilt as a fractional part of the power of $\delta$ and $\varepsilon$ actually allows one to expand in (\ref{grad_exp}) the $\delta$ and $\varepsilon$ corrections,
	\begin{align}
    &\delta^{3+8\frac{n_s-1}3}\!\simeq
    \Bigl(\frac{-X}{\gamma_*}\Bigr)^{\frac32}\!
    +4\,\frac{n_s-1}3\,\Bigl(\frac{-X}{\gamma_*}\Bigr)^{\frac32}\!
    \ln\!\frac{-X}{\gamma_*},\\
    &\varepsilon^{3+ 8\frac{n_s-1}3}\!\simeq
    \Bigl(\frac{3H_0\phi}{\sqrt{\gamma_*}}\Bigr)^3\!\!
    +8\frac{n_s-1}3\Bigl(\frac{3H_0\phi}{\sqrt{\gamma_*}}\Bigr)^3
    \ln\!\frac{3H_0\phi}{\sqrt{\gamma_*}}.
	\end{align}
Then the $\varepsilon$-terms in (\ref{grad_exp}) strongly resemble slow varying logarithmic quantum corrections of the Coleman-Weinberg type, whereas $\delta$-terms represent a gradient expansion of quantum effective action. This suggests the hypothesis that both versions of the GUMG inflation -- original constrained and the k-essence type ones -- are the effective theory of some fundamental quantum model.

Of course, this hypothesis requires justification from the viewpoint of known restrictions imposed by unitarity on the coefficients of the gradient expansion of the effective action (see for example \cite{positivity}), but this observation serves as a sufficiently strong motivation for rather special choice of the function (\ref{N_of_gamma_CMB}) and its $(1-n_s)$-expansion, which otherwise could be considered somewhat contrived. Together with an interesting naturalness property of this model (the mechanism of generating the needed level of tensor-to-scalar ratio $r\sim (e^{N})^{-4(1-n_s)}\sim0.001$ without invoking exponentially small or big parameters except the phenomenological estimate for e-folding number $N\sim 60$ \cite{GUMG-infl}) this observation provides a firmer ground for GUMG inflation model.

\section{Discussion}
We have established a nontrivial dynamical equivalence between generalized unimodular gravity (GUMG) and a particular class of k-essence theories which correspond to relativistic superfluids \cite{Greiter:1989qb,Son:2000ht,Son:2002zn}.Thus, GUMG is a novel way of describing relativistic hydrodynamics. We first covariantized the initially non-covariant GUMG theory by introducing coordinate Stueckelberg fields $\phi^A=(\phi^0,\phi^a)$. Then we observed that the spatial Stueckelberg fields $\phi^a$ completely decouple from the dynamics of the metric and temporal Stueckelberg field $\phi^0\equiv\phi$. This decoupling fully matches with the existence of four formally conserved charges $Q_A$ associated with the shift symmetry of Stueckelberg formulation, which actually degenerate to just one temporal charge $Q_0$. The k-essence Lagrangian of the field $\phi$ can be reconstructed from the the metric constraint of the original GUMG theory.
The parameters of cosmological perturbations in GUMG theory (in particular, their nontrivial speed of sound in the scalar sector) are explained from the viewpoint of hydrodynamical perfect fluid dynamics of k-essence field $\phi$. The previously derived model of GUMG inflation \cite{GUMG-infl}, when cast into the language of k-inflation, has the action in the form of gradient and field expansion strongly resembling effective field theory description.

The obtained dynamical equivalence is rather peculiar. Lagrangian equations of motion derived from both formulations turn out to be equivalent in the k-essence-metric sector, whereas spatial Stueckelberg fields fully decouple. At the same time it is impossible to say that k-essence formulation is the result of integrating out a subset of fields -- the Lagrange multiplier and these spatial Stueckelberg fields. This is because the Lagrange multiplier enters the action linearly and its elimination requires to solve the equations from the complementary subset -- a part of Einstein equations. In given examples of the reconstruction procedure this is the use of the variational equation for the metric scale factor, whose first integral yields the constant of motion $C$ (or equivalently $H_0$). As a result initial conditions enter the k-essence Lagrangian, which is somewhat unusual, but not contradictory\footnote{Physical boundary conditions always implicitly enter the action of the theory like, for example, asymptotically flat boundary conditions in Einstein-Hilbert action with the Gibbons-Hawking term. In effective field theories integration of certain fields out also brings into the action the dependence on boundary or initial conditions, but this ``weak'' dependence is usually disregarded.}.

As mentioned above, the covariantization procedure in constrained gravity models can be associated with a large variety of self-gravitating media theories. It should be emphasized, however, that there is a big difference from these theories in which noncovariant terms supporting $SO(3)$ or other types of global symmetry are explicitly introduced into the action. Contrary to this, in constrained models the starting point is a covariant gravity theory (which itself can be a generalization of Einstein gravity) in which metric coefficients are subject to a noncovariant constraint which, being introduced into the action with the Lagrange multiplier, generates effective matter stress tensor. Therefore in constrained models Stueckelberg covariantization is followed by decoupling and reconstruction---conversion of the Lagrange multiplier into the function of k-essence field. Such steps are absent in the formalism of self-gravitating theories.

Equivalence to perfect fluids can be extended beyond GUMG to a wider class of apparently modified gravity models with constraints of the form (\ref{W_constr}) or (\ref{W_constr_cov}), which are also characterized by perfect fluid type effective stress tensor.  Like in UMG and GUMG theories their Stueckelberg covariantization and reconstruction of an equivalent Lagrangian is also possible, but the resulting theory is a multiple fields one -- three spatial Stueckelberg fields remain gravitating, while the temporal one gets decoupled. Hence, contrary to k-essence, the equivalent perfect fluid possesses nonvanishing vorticity. This action describes fluids going beyond those considered in \cite{self-gravitating-fluid,Dubovsky:2005xd,Andersson:2020uue}. This case is briefly considered in Appendix \ref{app}.

Furthermore, the restriction to metric constraints corresponding to perfect fluid stress tensors, made in this work, can be relaxed without changing the main result---decoupling of the part of Stueckelberg fields. It is easy to check that the temporal Stueckelberg field decouples not only for the constraints of the form (\ref{W_constr}), but also for the class of constraints with arbitrary dependence on spatial part of contravariant metric coefficients $g^{ij}$. The $O(3)$-invariant subclass  of such theories (in which the constraint depends on $O(3)$-invariants $\tau_n = \tr \boldsymbol g^n$, $n=1,2,3$, $\boldsymbol g$ being the matrix composed of $g^{ij}$) after this decoupling corresponds to the model of so-called ``solid'' \cite{self-gravitating-infl} or ``elastic'' \cite{Gruzinov:2004ty} inflation.

A possible generalization to theories with matter sources is also possible, the effect of matter fields bringing interesting modification to the reconstruction procedure. Similarly to the above, inclusion of matter can be explicitly done for homogeneous Friedmann metric ansatz. Indeed, in this case the integrand in (\ref{phi_of_a}) acquires additional factor $(1 + \rho_m / \rho_\text{GUMG})^{-1/2}$. Here $\rho_m$ is the energy density of matter (which evolves as $\rho_m \propto a^{-3(1+w_m)}$ with $w_m$ being a barotropic parameter of matter), whereas $\rho_\text{GUMG} = C P(X)$ is the effective energy density of GUMG k-essence. When substituted to (\ref{K_def}), it changes the shape of $K(\phi)$ at the matter-dominated stage. This phenomenon can be referred to as tracking behavior of the theory.

It is well known that perfect fluids develop caustics and shock waves. In particular, for k-essence this issue was recently discussed in \cite{Babichev:2016hys,Frolov:2002rr}. The equivalence demonstrated in our work shows that, contrary to UMG, the GUMG is also vulnerable to these singular regimes. However, there are sigma-model like theories which can approximate, or in some sense UV complete, k-essence and which are free from these problems, see e.g. \cite{Bilic:2008zz,Bilic:2008zk,Tolley:2009fg,Babichev:2017lrx,Babichev:2018twg,Mukohyama:2020lsu}. It is very interesting to investigate whether due to the demonstrated here dynamical equivalence such sigma-model like theories can UV complete GUMG. This completion would also allow to use more standard methods to address questions like positivity of energy, similarly to how it is done for k-essence in \cite{positivity}. Finally, it is important to investigate whether the equivalence can be extended beyond classical dynamics to quantum realm.

\section*{Acknowledgments} We want to express special thanks to Valery Rubakov for helpful comments. N.~K. and A.~V. are grateful to the organisers of the Moscow International School of Physics, and in particular to Mikhail Danilov, for their warm hospitality, during the very first stages of this work. It is also a pleasure to thank Riccardo Rattazzi for useful discussions which partially motivated this study. This work was supported by the RFBR grant No.20-02-00297 and by the Foundation for Theoretical Physics Development “Basis”. The work of A.~V. is supported by the J.~E.~Purkyn\v{e} Fellowship of the Czech Academy of Sciences and by the Grant Agency of the Czech Republic, GA\v{C}R grant 20-28525S.

\appendix
\renewcommand{\thesection}{\Alph{section}}
\renewcommand{\theequation}{\Alph{section}.\arabic{equation}}

\section{Stueckelberg fields decoupling for the constraint family (\ref{W_constr_cov})\label{app}}
	
	Here we consider the second class of modified gravity theories (\ref{action_mod_noncov}) with $SO(3)$-invariant effective stress tensor of perfect fluid. Its constraint function $U(g^{\mu\nu})$ has a special metric dependence (\ref{W_constr}) which takes the form (\ref{W_constr_cov}) after covariantization procedure (\ref{cov_rule}). We assume that this constraint can be solved with respect to $Z$ as $Z = Q(b)$ and set in (\ref{action_mod_noncov}) $U = Q(b) - Z$. Thus, the action takes the form
	\begin{equation}
	S[g_{\mu\nu}, \Lambda, \phi^A] = S_\text{EH}[g_{\mu\nu}]
    + \int d^4x \sqrt{-g} \, \Lambda \, (Q(b) - Z), \label{action_pf}
	\end{equation}
where as before $Z = \sqrt{-\det C^{AB}}$ and $b = \sqrt{\det C^{ab}}$.
The corresponding metric stress tensor reads
	\begin{eqnarray}
	&&T_{\mu\nu} = \Lambda \, \bigl[(Q - b \, Q_b) \, g_{\mu\nu}
    - b \, Q_b \, u_\mu u_\nu\bigr],\\
    &&u^\mu = \frac1{3!\sqrt{-g} \, b} \epsilon^{\mu\nu\rho\sigma} \epsilon_{abc} \partial_\nu \phi^a \partial_\rho \phi^b \partial_\sigma \phi^c,
	\end{eqnarray}
where $Q_b=\partial_b Q$. Thus, contrary to k-essence, the equivalent perfect fluid possesses nonvanishing vorticity, see \cite{Dubovsky:2005xd}. Its hydrodynamical parameters are
	\begin{equation}
	p = \Lambda \, (Q - b \, Q_b) = w \rho, \quad
    w = -1 + \frac{b \, Q_b}{Q}. \label{set_fluid}
	\end{equation}

As in the case (\ref{action_GUMG_cov}), corresponding to the generalized unimodular gravity, part of Stueckelberg fields decouples from the theory's dynamics. Indeed, the stress tensor (\ref{set_fluid}) does not depend on the temporal Stueckelberg field $\phi^0$, so that $\phi^0$ does not participate in Einstein equations. As it was shown in Section \ref{sec_cov}, the Stueckelberg field equations do not introduce new dynamics and follow from the Einstein equations. The only place that involves temporal Stueckelberg field is the constraint $Z = Q(b)$, which expresses temporal Stueckelberg fields in terms of spatial ones. In that sense, $\phi^0$ decouples from the dynamics of the theory.
	
	In the present case, there is an analogue of the conservation law (\ref{cons_proj}). Projecting the conservation law $\nabla^\mu T_{\mu\nu} = 0$ of stress-energy tensor (\ref{set_fluid}) onto velocity vector $u^\nu$, one gets
	\begin{equation}
	u^\nu \nabla^\mu T_{\mu\nu} = Q(b) \, u^\mu \partial_\mu \Lambda = 0.
	\end{equation}
This relation implies that $\Lambda = L(\boldsymbol\phi)$ is some function of spatial Stueckelberg fields $\boldsymbol\phi\equiv(\phi^1,\phi^2,\phi^3)$, since $u^\mu \partial_\mu \phi^a = 0$. Thus, one allows to rewrite spatial Stueckelberg field equations as
	\begin{equation}
	\frac1{\sqrt{-g}} \frac{\delta S}{\delta \phi^a}
    = \left(\frac{\partial}{\partial\phi^a} - \nabla_\mu \frac{\partial}{\partial (\nabla_\mu \phi^a)}\right) L(\boldsymbol\phi) Q(b) = 0,
	\end{equation}
whereas stress tensor takes the form of
	\begin{equation}
	T_{\mu\nu} = L(\boldsymbol\phi) \left[
    \bigl(Q(b) - b \, Q_b(b)\bigr) \, g_{\mu\nu} - b \, Q_b (b) \, u_\mu u_\nu\right]. \label{set_pf}
	\end{equation}

These field equations and stress tensor can be generated by the following scalar field theory, minimally coupled to the Einstein gravity
	\begin{equation}
	S_\mathrm{pf}[g_{\mu\nu}, \phi^a] = S_\text{EH}[g_{\mu\nu}]
    + \int d^4x \sqrt{-g} \,
    L(\boldsymbol\phi)\, Q(b). \label{action_pf_eq}
	\end{equation}
As in the case of (\ref{action_GUMG_cov}), the unknown function $L(\boldsymbol\phi)$ should be found from the full set of Einstein equations with the stress tensor (\ref{set_pf}), which can be done in quadratures only on homogeneous backgrounds.

Unfortunately, the case of Friedmann background is not very indicative of the procedure, because the only choice of $\phi^a$ compatible with homogeneity of the stress tensor is again $\phi^a=x^a=x^i$, and the procedure trivializes.
Indeed, on Friedmannian background (\ref{bg_fr}) $Z=\dot\phi/{\cal N}a^3$ and $b=1/a^3$, so that the action (\ref{action_pf}) reads
	\begin{align}
	S[a, \mathcal N, \phi, \Lambda] = &-3 \int dt \, \mathcal N a^3
    \frac{\dot a^2}{\mathcal N^2 a^2}\nonumber\\
    & + \int dt \, \Lambda \, (\mathcal N a^3 \, Q(b) - \dot \phi).
	\end{align}
Its variation with respect to $\phi$ gives $\Lambda$ as a constant in time, which is determined by initial conditions. The assumption of spatial homogeneity then implies that this is also a constant in space, $L(\boldsymbol\phi) = \Lambda = \text{const}$. Cosmological applications of such theories are well studied \cite{self-gravitating-fluid} and, in particular, include generation of primordial power spectrum consistent with current CMB data \cite{self-gravitating-infl}.

A homogeneous background nontrivially employing the constraint $U = Q(b) - Z=0$ is the metric with nonvanishing spatially constant shift functions $ds^2 = - \mathcal N^2(t)\, dt^2 + a^2(t)\, \delta_{ij} dx^i dx^j+2{\cal N}_i(t)dx^i\,dt$. These shift functions take up the role of dynamically inert $\phi^i=x^i$ and realize the principal difference of the constraint (\ref{W_constr}) from (\ref{V_constr}) mentioned in Introduction. In contrast to (\ref{V_constr}) the constraint (\ref{W_constr}) implicitly expresses the lapse function not only in terms of $\gamma_{ij}$ but also as functions of ${\cal N}_i$, ${\cal N}=N(\gamma_{ij},{\cal N}_i)$. In the covariantized form it is obvious from the expressions which are valid on Friedmann ansatz (with $\phi^i=x^i$)
    \begin{equation}
	Z= \frac{\dot\phi}{\mathcal N a^3},\quad
    b=\frac1{a^3}\,\sqrt{1-
    \frac{\delta^{ij}{\cal N}_i{\cal N}_j}{a^2{\cal N}^2}},
	\end{equation}
so that the constraint $Q(b)-Z=0$ establishes additional dependence of $\cal N$ on ${\cal N}_i$.
	
\bibliography{article_arxiv}

\begin{thebibliography}{55}%
\makeatletter
\providecommand \@ifxundefined [1]{%
 \@ifx{#1\undefined}
}%
\providecommand \@ifnum [1]{%
 \ifnum #1\expandafter \@firstoftwo
 \else \expandafter \@secondoftwo
 \fi
}%
\providecommand \@ifx [1]{%
 \ifx #1\expandafter \@firstoftwo
 \else \expandafter \@secondoftwo
 \fi
}%
\providecommand \natexlab [1]{#1}%
\providecommand \enquote  [1]{``#1''}%
\providecommand \bibnamefont  [1]{#1}%
\providecommand \bibfnamefont [1]{#1}%
\providecommand \citenamefont [1]{#1}%
\providecommand \href@noop [0]{\@secondoftwo}%
\providecommand \href [0]{\begingroup \@sanitize@url \@href}%
\providecommand \@href[1]{\@@startlink{#1}\@@href}%
\providecommand \@@href[1]{\endgroup#1\@@endlink}%
\providecommand \@sanitize@url [0]{\catcode `\\12\catcode `\$12\catcode
  `\&12\catcode `\#12\catcode `\^12\catcode `\_12\catcode `\%12\relax}%
\providecommand \@@startlink[1]{}%
\providecommand \@@endlink[0]{}%
\providecommand \url  [0]{\begingroup\@sanitize@url \@url }%
\providecommand \@url [1]{\endgroup\@href {#1}{\urlprefix }}%
\providecommand \urlprefix  [0]{URL }%
\providecommand \Eprint [0]{\href }%
\providecommand \doibase [0]{http://dx.doi.org/}%
\providecommand \selectlanguage [0]{\@gobble}%
\providecommand \bibinfo  [0]{\@secondoftwo}%
\providecommand \bibfield  [0]{\@secondoftwo}%
\providecommand \translation [1]{[#1]}%
\providecommand \BibitemOpen [0]{}%
\providecommand \bibitemStop [0]{}%
\providecommand \bibitemNoStop [0]{.\EOS\space}%
\providecommand \EOS [0]{\spacefactor3000\relax}%
\providecommand \BibitemShut  [1]{\csname bibitem#1\endcsname}%
\let\auto@bib@innerbib\@empty
\bibitem [{\citenamefont {{Starobinsky}}(1980)}]{Starobinsky}%
  \BibitemOpen
  \bibfield  {author} {\bibinfo {author} {\bibfnamefont {A.~A.}\ \bibnamefont
  {{Starobinsky}}},\ }\bibfield  {title} {\enquote {\bibinfo {title} {A new
  type of isotropic cosmological models without singularity},}\ }\href
  {\doibase 10.1016/0370-2693(80)90670-X} {\bibfield  {journal} {\bibinfo
  {journal} {Phys. Lett. B}\ }\textbf {\bibinfo {volume} {91}},\ \bibinfo
  {pages} {99--102} (\bibinfo {year} {1980})}\BibitemShut {NoStop}%
\bibitem [{\citenamefont {Whitt}(1984)}]{Whitt:1984pd}%
  \BibitemOpen
  \bibfield  {author} {\bibinfo {author} {\bibfnamefont {Brian}\ \bibnamefont
  {Whitt}},\ }\bibfield  {title} {\enquote {\bibinfo {title} {Fourth order
  gravity as general relativity plus matter},}\ }\href {\doibase
  10.1016/0370-2693(84)90332-0} {\bibfield  {journal} {\bibinfo  {journal}
  {Phys. Lett. B}\ }\textbf {\bibinfo {volume} {145}},\ \bibinfo {pages}
  {176--178} (\bibinfo {year} {1984})}\BibitemShut {NoStop}%
\bibitem [{\citenamefont {Schmidt}(1987)}]{Schmidt:2001ac}%
  \BibitemOpen
  \bibfield  {author} {\bibinfo {author} {\bibfnamefont {Hans-Juergen}\
  \bibnamefont {Schmidt}},\ }\bibfield  {title} {\enquote {\bibinfo {title}
  {{Comparing selfinteracting scalar fields and $R + R^3$ cosmological
  models}},}\ }\href@noop {} {\bibfield  {journal} {\bibinfo  {journal}
  {Astron. Nachr.}\ }\textbf {\bibinfo {volume} {308}},\ \bibinfo {pages}
  {183--188} (\bibinfo {year} {1987})},\ \Eprint
  {http://arxiv.org/abs/gr-qc/0106035} {arXiv:gr-qc/0106035} \BibitemShut
  {NoStop}%
\bibitem [{\citenamefont {van~der Bij}\ \emph {et~al.}(1982)\citenamefont
  {van~der Bij}, \citenamefont {van Dam},\ and\ \citenamefont
  {Ng}}]{vanderBij:1981ym}%
  \BibitemOpen
  \bibfield  {author} {\bibinfo {author} {\bibfnamefont {J.~J.}\ \bibnamefont
  {van~der Bij}}, \bibinfo {author} {\bibfnamefont {H.}~\bibnamefont {van
  Dam}}, \ and\ \bibinfo {author} {\bibfnamefont {Yee~Jack}\ \bibnamefont
  {Ng}},\ }\bibfield  {title} {\enquote {\bibinfo {title} {The exchange of
  massless spin two particles},}\ }\href {\doibase
  10.1016/0378-4371(82)90247-3} {\bibfield  {journal} {\bibinfo  {journal}
  {Physica}\ }\textbf {\bibinfo {volume} {116A}},\ \bibinfo {pages} {307--320}
  (\bibinfo {year} {1982})}\BibitemShut {NoStop}%
\bibitem [{\citenamefont {Buchmuller}\ and\ \citenamefont
  {Dragon}(1988)}]{Buchmuller:1988wx}%
  \BibitemOpen
  \bibfield  {author} {\bibinfo {author} {\bibfnamefont {W.}~\bibnamefont
  {Buchmuller}}\ and\ \bibinfo {author} {\bibfnamefont {N.}~\bibnamefont
  {Dragon}},\ }\bibfield  {title} {\enquote {\bibinfo {title} {Einstein gravity
  from restricted coordinate invariance},}\ }\href {\doibase
  10.1016/0370-2693(88)90577-1} {\bibfield  {journal} {\bibinfo  {journal}
  {Phys. Lett.}\ }\textbf {\bibinfo {volume} {B207}},\ \bibinfo {pages}
  {292--294} (\bibinfo {year} {1988})}\BibitemShut {NoStop}%
\bibitem [{\citenamefont {Henneaux}\ and\ \citenamefont
  {Teitelboim}(1989)}]{UMG_HT}%
  \BibitemOpen
  \bibfield  {author} {\bibinfo {author} {\bibfnamefont {M.}~\bibnamefont
  {Henneaux}}\ and\ \bibinfo {author} {\bibfnamefont {C.}~\bibnamefont
  {Teitelboim}},\ }\bibfield  {title} {\enquote {\bibinfo {title} {{The
  Cosmological Constant and General Covariance}},}\ }\href {\doibase
  10.1016/0370-2693(89)91251-3} {\bibfield  {journal} {\bibinfo  {journal}
  {Phys. Lett. B}\ }\textbf {\bibinfo {volume} {222}},\ \bibinfo {pages}
  {195--199} (\bibinfo {year} {1989})}\BibitemShut {NoStop}%
\bibitem [{\citenamefont {Unruh}(1989)}]{UMG_Unruh}%
  \BibitemOpen
  \bibfield  {author} {\bibinfo {author} {\bibfnamefont {W.G.}\ \bibnamefont
  {Unruh}},\ }\bibfield  {title} {\enquote {\bibinfo {title} {{A Unimodular
  Theory of Canonical Quantum Gravity}},}\ }\href {\doibase
  10.1103/PhysRevD.40.1048} {\bibfield  {journal} {\bibinfo  {journal} {Phys.
  Rev. D}\ }\textbf {\bibinfo {volume} {40}},\ \bibinfo {pages} {1048}
  (\bibinfo {year} {1989})}\BibitemShut {NoStop}%
\bibitem [{\citenamefont {Horava}(2009)}]{Horava}%
  \BibitemOpen
  \bibfield  {author} {\bibinfo {author} {\bibfnamefont {Petr}\ \bibnamefont
  {Horava}},\ }\bibfield  {title} {\enquote {\bibinfo {title} {{Quantum Gravity
  at a Lifshitz Point}},}\ }\href {\doibase 10.1103/PhysRevD.79.084008}
  {\bibfield  {journal} {\bibinfo  {journal} {Phys. Rev. D}\ }\textbf {\bibinfo
  {volume} {79}},\ \bibinfo {pages} {084008} (\bibinfo {year} {2009})},\
  \Eprint {http://arxiv.org/abs/0901.3775} {arXiv:0901.3775 [hep-th]}
  \BibitemShut {NoStop}%
\bibitem [{\citenamefont {Fierz}\ and\ \citenamefont
  {Pauli}(1939)}]{massive_fp}%
  \BibitemOpen
  \bibfield  {author} {\bibinfo {author} {\bibfnamefont {M.}~\bibnamefont
  {Fierz}}\ and\ \bibinfo {author} {\bibfnamefont {W.}~\bibnamefont {Pauli}},\
  }\bibfield  {title} {\enquote {\bibinfo {title} {{On relativistic wave
  equations for particles of arbitrary spin in an electromagnetic field}},}\
  }\href {\doibase 10.1098/rspa.1939.0140} {\bibfield  {journal} {\bibinfo
  {journal} {Proc. Roy. Soc. Lond. A}\ }\textbf {\bibinfo {volume} {173}},\
  \bibinfo {pages} {211--232} (\bibinfo {year} {1939})}\BibitemShut {NoStop}%
\bibitem [{\citenamefont {Rubakov}\ and\ \citenamefont
  {Tinyakov}(2008)}]{Rubakov:2008nh}%
  \BibitemOpen
  \bibfield  {author} {\bibinfo {author} {\bibfnamefont {V.A.}\ \bibnamefont
  {Rubakov}}\ and\ \bibinfo {author} {\bibfnamefont {P.G.}\ \bibnamefont
  {Tinyakov}},\ }\bibfield  {title} {\enquote {\bibinfo {title}
  {Infrared-modified gravities and massive gravitons},}\ }\href {\doibase
  10.1070/PU2008v051n08ABEH006600} {\bibfield  {journal} {\bibinfo  {journal}
  {Phys. Usp.}\ }\textbf {\bibinfo {volume} {51}},\ \bibinfo {pages} {759--792}
  (\bibinfo {year} {2008})},\ \Eprint {http://arxiv.org/abs/0802.4379}
  {arXiv:0802.4379 [hep-th]} \BibitemShut {NoStop}%
\bibitem [{\citenamefont {de~Rham}(2014)}]{deRham:2014zqa}%
  \BibitemOpen
  \bibfield  {author} {\bibinfo {author} {\bibfnamefont {Claudia}\ \bibnamefont
  {de~Rham}},\ }\bibfield  {title} {\enquote {\bibinfo {title} {Massive
  gravity},}\ }\href {\doibase 10.12942/lrr-2014-7} {\bibfield  {journal}
  {\bibinfo  {journal} {Living Rev. Rel.}\ }\textbf {\bibinfo {volume} {17}},\
  \bibinfo {pages} {7} (\bibinfo {year} {2014})},\ \Eprint
  {http://arxiv.org/abs/1401.4173} {arXiv:1401.4173 [hep-th]} \BibitemShut
  {NoStop}%
\bibitem [{\citenamefont {Lin}\ and\ \citenamefont
  {Mukohyama}(2017)}]{Lin:2017oow}%
  \BibitemOpen
  \bibfield  {author} {\bibinfo {author} {\bibfnamefont {Chunshan}\
  \bibnamefont {Lin}}\ and\ \bibinfo {author} {\bibfnamefont {Shinji}\
  \bibnamefont {Mukohyama}},\ }\bibfield  {title} {\enquote {\bibinfo {title}
  {A class of minimally modified gravity theories},}\ }\href {\doibase
  10.1088/1475-7516/2017/10/033} {\bibfield  {journal} {\bibinfo  {journal}
  {JCAP}\ }\textbf {\bibinfo {volume} {10}},\ \bibinfo {pages} {033} (\bibinfo
  {year} {2017})},\ \Eprint {http://arxiv.org/abs/1708.03757} {arXiv:1708.03757
  [gr-qc]} \BibitemShut {NoStop}%
\bibitem [{\citenamefont {Gao}(2014)}]{Gao:2014soa}%
  \BibitemOpen
  \bibfield  {author} {\bibinfo {author} {\bibfnamefont {Xian}\ \bibnamefont
  {Gao}},\ }\bibfield  {title} {\enquote {\bibinfo {title} {Unifying framework
  for scalar-tensor theories of gravity},}\ }\href {\doibase
  10.1103/PhysRevD.90.081501} {\bibfield  {journal} {\bibinfo  {journal} {Phys.
  Rev. D}\ }\textbf {\bibinfo {volume} {90}},\ \bibinfo {pages} {081501}
  (\bibinfo {year} {2014})},\ \Eprint {http://arxiv.org/abs/1406.0822}
  {arXiv:1406.0822 [gr-qc]} \BibitemShut {NoStop}%
\bibitem [{\citenamefont {Eling}\ \emph {et~al.}(2004)\citenamefont {Eling},
  \citenamefont {Jacobson},\ and\ \citenamefont {Mattingly}}]{aether}%
  \BibitemOpen
  \bibfield  {author} {\bibinfo {author} {\bibfnamefont {Christopher}\
  \bibnamefont {Eling}}, \bibinfo {author} {\bibfnamefont {Ted}\ \bibnamefont
  {Jacobson}}, \ and\ \bibinfo {author} {\bibfnamefont {David}\ \bibnamefont
  {Mattingly}},\ }\bibfield  {title} {\enquote {\bibinfo {title}
  {{Einstein-Aether theory}},}\ }in\ \href@noop {} {\emph {\bibinfo {booktitle}
  {{Deserfest: A Celebration of the Life and Works of Stanley Deser}}}}\
  (\bibinfo {year} {2004})\ pp.\ \bibinfo {pages} {163--179},\ \Eprint
  {http://arxiv.org/abs/gr-qc/0410001} {arXiv:gr-qc/0410001} \BibitemShut
  {NoStop}%
\bibitem [{\citenamefont {Ballesteros}\ \emph {et~al.}(2016)\citenamefont
  {Ballesteros}, \citenamefont {Comelli},\ and\ \citenamefont
  {Pilo}}]{self-gravitating}%
  \BibitemOpen
  \bibfield  {author} {\bibinfo {author} {\bibfnamefont {Guillermo}\
  \bibnamefont {Ballesteros}}, \bibinfo {author} {\bibfnamefont {Denis}\
  \bibnamefont {Comelli}}, \ and\ \bibinfo {author} {\bibfnamefont {Luigi}\
  \bibnamefont {Pilo}},\ }\bibfield  {title} {\enquote {\bibinfo {title}
  {{Massive and modified gravity as self-gravitating media}},}\ }\href
  {\doibase 10.1103/PhysRevD.94.124023} {\bibfield  {journal} {\bibinfo
  {journal} {Phys. Rev. D}\ }\textbf {\bibinfo {volume} {94}},\ \bibinfo
  {pages} {124023} (\bibinfo {year} {2016})},\ \Eprint
  {http://arxiv.org/abs/1603.02956} {arXiv:1603.02956 [hep-th]} \BibitemShut
  {NoStop}%
\bibitem [{\citenamefont {Arnowitt}\ \emph {et~al.}(2008)\citenamefont
  {Arnowitt}, \citenamefont {Deser},\ and\ \citenamefont
  {Misner}}]{Arnowitt:1962hi}%
  \BibitemOpen
  \bibfield  {author} {\bibinfo {author} {\bibfnamefont {Richard~L.}\
  \bibnamefont {Arnowitt}}, \bibinfo {author} {\bibfnamefont {Stanley}\
  \bibnamefont {Deser}}, \ and\ \bibinfo {author} {\bibfnamefont {Charles~W.}\
  \bibnamefont {Misner}},\ }\bibfield  {title} {\enquote {\bibinfo {title} {The
  dynamics of general relativity},}\ }\href {\doibase
  10.1007/s10714-008-0661-1} {\bibfield  {journal} {\bibinfo  {journal} {Gen.
  Rel. Grav.}\ }\textbf {\bibinfo {volume} {40}},\ \bibinfo {pages}
  {1997--2027} (\bibinfo {year} {2008})},\ \Eprint
  {http://arxiv.org/abs/gr-qc/0405109} {arXiv:gr-qc/0405109 [gr-qc]}
  \BibitemShut {NoStop}%
\bibitem [{\citenamefont {Poisson}(2004)}]{Poisson}%
  \BibitemOpen
  \bibfield  {author} {\bibinfo {author} {\bibfnamefont {Eric}\ \bibnamefont
  {Poisson}},\ }\href@noop {} {\emph {\bibinfo {title} {A Relativist's Toolkit,
  The Mathematics of Black-Hole Mechanics}}}\ (\bibinfo  {publisher} {Cambridge
  University Press},\ \bibinfo {year} {2004})\BibitemShut {NoStop}%
\bibitem [{\citenamefont {Barvinsky}\ and\ \citenamefont
  {Kamenshchik}(2017)}]{GUMG}%
  \BibitemOpen
  \bibfield  {author} {\bibinfo {author} {\bibfnamefont {A.O.}\ \bibnamefont
  {Barvinsky}}\ and\ \bibinfo {author} {\bibfnamefont {A.~Yu.}\ \bibnamefont
  {Kamenshchik}},\ }\bibfield  {title} {\enquote {\bibinfo {title} {{Darkness
  without dark matter and energy -- generalized unimodular gravity}},}\ }\href
  {\doibase 10.1016/j.physletb.2017.09.045} {\bibfield  {journal} {\bibinfo
  {journal} {Phys. Lett. B}\ }\textbf {\bibinfo {volume} {774}},\ \bibinfo
  {pages} {59--63} (\bibinfo {year} {2017})},\ \Eprint
  {http://arxiv.org/abs/1705.09470} {arXiv:1705.09470 [gr-qc]} \BibitemShut
  {NoStop}%
\bibitem [{\citenamefont {Barvinsky}\ \emph {et~al.}(2019)\citenamefont
  {Barvinsky}, \citenamefont {Kolganov}, \citenamefont {Kurov},\ and\
  \citenamefont {Nesterov}}]{GUMG-can}%
  \BibitemOpen
  \bibfield  {author} {\bibinfo {author} {\bibfnamefont {A.O.}\ \bibnamefont
  {Barvinsky}}, \bibinfo {author} {\bibfnamefont {N.}~\bibnamefont {Kolganov}},
  \bibinfo {author} {\bibfnamefont {A.}~\bibnamefont {Kurov}}, \ and\ \bibinfo
  {author} {\bibfnamefont {D.}~\bibnamefont {Nesterov}},\ }\bibfield  {title}
  {\enquote {\bibinfo {title} {{Dynamics of the generalized unimodular gravity
  theory}},}\ }\href {\doibase 10.1103/PhysRevD.100.023542} {\bibfield
  {journal} {\bibinfo  {journal} {Phys. Rev. D}\ }\textbf {\bibinfo {volume}
  {100}},\ \bibinfo {pages} {023542} (\bibinfo {year} {2019})},\ \Eprint
  {http://arxiv.org/abs/1903.09897} {arXiv:1903.09897 [hep-th]} \BibitemShut
  {NoStop}%
\bibitem [{\citenamefont {Henneaux}\ and\ \citenamefont
  {Teitelboim}(1992)}]{HT-book}%
  \BibitemOpen
  \bibfield  {author} {\bibinfo {author} {\bibfnamefont {M.}~\bibnamefont
  {Henneaux}}\ and\ \bibinfo {author} {\bibfnamefont {C.}~\bibnamefont
  {Teitelboim}},\ }\href@noop {} {\emph {\bibinfo {title} {{Quantization of
  gauge systems}}}}\ (\bibinfo  {publisher} {Princeton University Press},\
  \bibinfo {address} {Princeton},\ \bibinfo {year} {1992})\BibitemShut
  {NoStop}%
\bibitem [{\citenamefont {Kaparulin}\ and\ \citenamefont
  {Lyakhovich}(2019{\natexlab{a}})}]{Lyakhovich}%
  \BibitemOpen
  \bibfield  {author} {\bibinfo {author} {\bibfnamefont {D.S.}\ \bibnamefont
  {Kaparulin}}\ and\ \bibinfo {author} {\bibfnamefont {S.L.}\ \bibnamefont
  {Lyakhovich}},\ }\bibfield  {title} {\enquote {\bibinfo {title} {{A note on
  unfree gauge symmetry}},}\ }\href {\doibase 10.1016/j.nuclphysb.2019.114735}
  {\bibfield  {journal} {\bibinfo  {journal} {Nucl. Phys. B}\ }\textbf
  {\bibinfo {volume} {947}},\ \bibinfo {pages} {114735} (\bibinfo {year}
  {2019}{\natexlab{a}})},\ \Eprint {http://arxiv.org/abs/1904.04038}
  {arXiv:1904.04038 [hep-th]} \BibitemShut {NoStop}%
\bibitem [{\citenamefont {Kaparulin}\ and\ \citenamefont
  {Lyakhovich}(2019{\natexlab{b}})}]{Lyakhovich_BV}%
  \BibitemOpen
  \bibfield  {author} {\bibinfo {author} {\bibfnamefont {D.S.}\ \bibnamefont
  {Kaparulin}}\ and\ \bibinfo {author} {\bibfnamefont {S.L.}\ \bibnamefont
  {Lyakhovich}},\ }\bibfield  {title} {\enquote {\bibinfo {title} {{Unfree
  gauge symmetry in the BV formalism}},}\ }\href {\doibase
  10.1140/epjc/s10052-019-7233-2} {\bibfield  {journal} {\bibinfo  {journal}
  {Eur. Phys. J. C}\ }\textbf {\bibinfo {volume} {79}},\ \bibinfo {pages} {718}
  (\bibinfo {year} {2019}{\natexlab{b}})},\ \Eprint
  {http://arxiv.org/abs/1907.03443} {arXiv:1907.03443 [hep-th]} \BibitemShut
  {NoStop}%
\bibitem [{\citenamefont {Barvinsky}\ and\ \citenamefont
  {Kolganov}(2019)}]{GUMG-infl}%
  \BibitemOpen
  \bibfield  {author} {\bibinfo {author} {\bibfnamefont {A.O.}\ \bibnamefont
  {Barvinsky}}\ and\ \bibinfo {author} {\bibfnamefont {N.}~\bibnamefont
  {Kolganov}},\ }\bibfield  {title} {\enquote {\bibinfo {title} {{Inflation in
  generalized unimodular gravity}},}\ }\href {\doibase
  10.1103/PhysRevD.100.123510} {\bibfield  {journal} {\bibinfo  {journal}
  {Phys. Rev. D}\ }\textbf {\bibinfo {volume} {100}},\ \bibinfo {pages}
  {123510} (\bibinfo {year} {2019})},\ \Eprint
  {http://arxiv.org/abs/1908.05697} {arXiv:1908.05697 [gr-qc]} \BibitemShut
  {NoStop}%
\bibitem [{\citenamefont {Barvinsky}\ \emph {et~al.}(2016)\citenamefont
  {Barvinsky}, \citenamefont {Blas}, \citenamefont {Herrero-Valea},
  \citenamefont {Sibiryakov},\ and\ \citenamefont {Steinwachs}}]{Horava-ren}%
  \BibitemOpen
  \bibfield  {author} {\bibinfo {author} {\bibfnamefont {Andrei~O.}\
  \bibnamefont {Barvinsky}}, \bibinfo {author} {\bibfnamefont {Diego}\
  \bibnamefont {Blas}}, \bibinfo {author} {\bibfnamefont {Mario}\ \bibnamefont
  {Herrero-Valea}}, \bibinfo {author} {\bibfnamefont {Sergey~M.}\ \bibnamefont
  {Sibiryakov}}, \ and\ \bibinfo {author} {\bibfnamefont {Christian~F.}\
  \bibnamefont {Steinwachs}},\ }\bibfield  {title} {\enquote {\bibinfo {title}
  {{Renormalization of Ho\v{r}ava gravity}},}\ }\href {\doibase
  10.1103/PhysRevD.93.064022} {\bibfield  {journal} {\bibinfo  {journal} {Phys.
  Rev. D}\ }\textbf {\bibinfo {volume} {93}},\ \bibinfo {pages} {064022}
  (\bibinfo {year} {2016})},\ \Eprint {http://arxiv.org/abs/1512.02250}
  {arXiv:1512.02250 [hep-th]} \BibitemShut {NoStop}%
\bibitem [{\citenamefont {Jirou\v{s}ek}\ and\ \citenamefont
  {Vikman}(2019)}]{Jirousek:2018ago}%
  \BibitemOpen
  \bibfield  {author} {\bibinfo {author} {\bibfnamefont {Pavel}\ \bibnamefont
  {Jirou\v{s}ek}}\ and\ \bibinfo {author} {\bibfnamefont {Alexander}\
  \bibnamefont {Vikman}},\ }\bibfield  {title} {\enquote {\bibinfo {title} {New
  weyl-invariant vector-tensor theory for the cosmological constant},}\ }\href
  {\doibase 10.1088/1475-7516/2019/04/004} {\bibfield  {journal} {\bibinfo
  {journal} {JCAP}\ }\textbf {\bibinfo {volume} {04}},\ \bibinfo {pages} {004}
  (\bibinfo {year} {2019})},\ \Eprint {http://arxiv.org/abs/1811.09547}
  {arXiv:1811.09547 [gr-qc]} \BibitemShut {NoStop}%
\bibitem [{\citenamefont {Kuchar}(1991)}]{UMG_Kuchar}%
  \BibitemOpen
  \bibfield  {author} {\bibinfo {author} {\bibfnamefont {Karel~V.}\
  \bibnamefont {Kuchar}},\ }\bibfield  {title} {\enquote {\bibinfo {title}
  {{Does an unspecified cosmological constant solve the problem of time in
  quantum gravity?}}}\ }\href {\doibase 10.1103/PhysRevD.43.3332} {\bibfield
  {journal} {\bibinfo  {journal} {Phys. Rev. D}\ }\textbf {\bibinfo {volume}
  {43}},\ \bibinfo {pages} {3332--3344} (\bibinfo {year} {1991})}\BibitemShut
  {NoStop}%
\bibitem [{\citenamefont {Hammer}\ \emph {et~al.}(2020)\citenamefont {Hammer},
  \citenamefont {Jirou\v{s}ek},\ and\ \citenamefont {Vikman}}]{UMG_Vikman}%
  \BibitemOpen
  \bibfield  {author} {\bibinfo {author} {\bibfnamefont {Katrin}\ \bibnamefont
  {Hammer}}, \bibinfo {author} {\bibfnamefont {Pavel}\ \bibnamefont
  {Jirou\v{s}ek}}, \ and\ \bibinfo {author} {\bibfnamefont {Alexander}\
  \bibnamefont {Vikman}},\ }\bibfield  {title} {\enquote {\bibinfo {title}
  {{Axionic cosmological constant}},}\ }\href@noop {} {\  (\bibinfo {year}
  {2020})},\ \Eprint {http://arxiv.org/abs/2001.03169} {arXiv:2001.03169
  [gr-qc]} \BibitemShut {NoStop}%
\bibitem [{\citenamefont {Blas}\ \emph {et~al.}(2009)\citenamefont {Blas},
  \citenamefont {Pujolas},\ and\ \citenamefont {Sibiryakov}}]{chronon}%
  \BibitemOpen
  \bibfield  {author} {\bibinfo {author} {\bibfnamefont {D.}~\bibnamefont
  {Blas}}, \bibinfo {author} {\bibfnamefont {O.}~\bibnamefont {Pujolas}}, \
  and\ \bibinfo {author} {\bibfnamefont {S.}~\bibnamefont {Sibiryakov}},\
  }\bibfield  {title} {\enquote {\bibinfo {title} {{On the Extra Mode and
  Inconsistency of Horava Gravity}},}\ }\href {\doibase
  10.1088/1126-6708/2009/10/029} {\bibfield  {journal} {\bibinfo  {journal}
  {JHEP}\ }\textbf {\bibinfo {volume} {10}},\ \bibinfo {pages} {029} (\bibinfo
  {year} {2009})},\ \Eprint {http://arxiv.org/abs/0906.3046} {arXiv:0906.3046
  [hep-th]} \BibitemShut {NoStop}%
\bibitem [{\citenamefont {Dubovsky}(2004)}]{massive_dubovsky}%
  \BibitemOpen
  \bibfield  {author} {\bibinfo {author} {\bibfnamefont {S.L.}\ \bibnamefont
  {Dubovsky}},\ }\bibfield  {title} {\enquote {\bibinfo {title} {{Phases of
  massive gravity}},}\ }\href {\doibase 10.1088/1126-6708/2004/10/076}
  {\bibfield  {journal} {\bibinfo  {journal} {JHEP}\ }\textbf {\bibinfo
  {volume} {10}},\ \bibinfo {pages} {076} (\bibinfo {year} {2004})},\ \Eprint
  {http://arxiv.org/abs/hep-th/0409124} {arXiv:hep-th/0409124} \BibitemShut
  {NoStop}%
\bibitem [{\citenamefont {Armendariz-Picon}\ \emph {et~al.}(1999)\citenamefont
  {Armendariz-Picon}, \citenamefont {Damour},\ and\ \citenamefont
  {Mukhanov}}]{k-infl}%
  \BibitemOpen
  \bibfield  {author} {\bibinfo {author} {\bibfnamefont {C.}~\bibnamefont
  {Armendariz-Picon}}, \bibinfo {author} {\bibfnamefont {T.}~\bibnamefont
  {Damour}}, \ and\ \bibinfo {author} {\bibfnamefont {Viatcheslav~F.}\
  \bibnamefont {Mukhanov}},\ }\bibfield  {title} {\enquote {\bibinfo {title}
  {{k - inflation}},}\ }\href {\doibase 10.1016/S0370-2693(99)00603-6}
  {\bibfield  {journal} {\bibinfo  {journal} {Phys. Lett. B}\ }\textbf
  {\bibinfo {volume} {458}},\ \bibinfo {pages} {209--218} (\bibinfo {year}
  {1999})},\ \Eprint {http://arxiv.org/abs/hep-th/9904075}
  {arXiv:hep-th/9904075} \BibitemShut {NoStop}%
\bibitem [{\citenamefont {Mukhanov}(2005)}]{Mukhanov_book}%
  \BibitemOpen
  \bibfield  {author} {\bibinfo {author} {\bibfnamefont {V.}~\bibnamefont
  {Mukhanov}},\ }\href@noop {} {\emph {\bibinfo {title} {{Physical Foundations
  of Cosmology}}}}\ (\bibinfo  {publisher} {Cambridge University Press},\
  \bibinfo {address} {Oxford},\ \bibinfo {year} {2005})\BibitemShut {NoStop}%
\bibitem [{\citenamefont {Garriga}\ and\ \citenamefont
  {Mukhanov}(1999)}]{k-pert}%
  \BibitemOpen
  \bibfield  {author} {\bibinfo {author} {\bibfnamefont {Jaume}\ \bibnamefont
  {Garriga}}\ and\ \bibinfo {author} {\bibfnamefont {Viatcheslav~F.}\
  \bibnamefont {Mukhanov}},\ }\bibfield  {title} {\enquote {\bibinfo {title}
  {{Perturbations in k-inflation}},}\ }\href {\doibase
  10.1016/S0370-2693(99)00602-4} {\bibfield  {journal} {\bibinfo  {journal}
  {Phys. Lett. B}\ }\textbf {\bibinfo {volume} {458}},\ \bibinfo {pages}
  {219--225} (\bibinfo {year} {1999})},\ \Eprint
  {http://arxiv.org/abs/hep-th/9904176} {arXiv:hep-th/9904176} \BibitemShut
  {NoStop}%
\bibitem [{\citenamefont {Landau}\ and\ \citenamefont
  {Lifshitz}(1980)}]{Landafshitz_Teorpol}%
  \BibitemOpen
  \bibfield  {author} {\bibinfo {author} {\bibfnamefont {L.~D.}\ \bibnamefont
  {Landau}}\ and\ \bibinfo {author} {\bibfnamefont {E.~M.}\ \bibnamefont
  {Lifshitz}},\ }\href@noop {} {\emph {\bibinfo {title} {Course of Theoretical
  Physics, Vol. 2, The Classical Theory of Fields}}},\ \bibinfo {edition}
  {{fourth english}}\ ed.\ (\bibinfo  {publisher} {Butterworth-Heinemann},\
  \bibinfo {year} {1980})\BibitemShut {NoStop}%
\bibitem [{\citenamefont {Afshordi}\ \emph {et~al.}(2007)\citenamefont
  {Afshordi}, \citenamefont {Chung},\ and\ \citenamefont
  {Geshnizjani}}]{cuscuton}%
  \BibitemOpen
  \bibfield  {author} {\bibinfo {author} {\bibfnamefont {Niayesh}\ \bibnamefont
  {Afshordi}}, \bibinfo {author} {\bibfnamefont {Daniel~J.H.}\ \bibnamefont
  {Chung}}, \ and\ \bibinfo {author} {\bibfnamefont {Ghazal}\ \bibnamefont
  {Geshnizjani}},\ }\bibfield  {title} {\enquote {\bibinfo {title} {{Cuscuton:
  A Causal Field Theory with an Infinite Speed of Sound}},}\ }\href {\doibase
  10.1103/PhysRevD.75.083513} {\bibfield  {journal} {\bibinfo  {journal} {Phys.
  Rev. D}\ }\textbf {\bibinfo {volume} {75}},\ \bibinfo {pages} {083513}
  (\bibinfo {year} {2007})},\ \Eprint {http://arxiv.org/abs/hep-th/0609150}
  {arXiv:hep-th/0609150} \BibitemShut {NoStop}%
\bibitem [{\citenamefont {Dvali}\ \emph {et~al.}(2000)\citenamefont {Dvali},
  \citenamefont {Gabadadze},\ and\ \citenamefont {Porrati}}]{DGP}%
  \BibitemOpen
  \bibfield  {author} {\bibinfo {author} {\bibfnamefont {G.R.}\ \bibnamefont
  {Dvali}}, \bibinfo {author} {\bibfnamefont {Gregory}\ \bibnamefont
  {Gabadadze}}, \ and\ \bibinfo {author} {\bibfnamefont {Massimo}\ \bibnamefont
  {Porrati}},\ }\bibfield  {title} {\enquote {\bibinfo {title} {{4-D gravity on
  a brane in 5-D Minkowski space}},}\ }\href {\doibase
  10.1016/S0370-2693(00)00669-9} {\bibfield  {journal} {\bibinfo  {journal}
  {Phys. Lett. B}\ }\textbf {\bibinfo {volume} {485}},\ \bibinfo {pages}
  {208--214} (\bibinfo {year} {2000})},\ \Eprint
  {http://arxiv.org/abs/hep-th/0005016} {arXiv:hep-th/0005016} \BibitemShut
  {NoStop}%
\bibitem [{\citenamefont {Mukhanov}\ and\ \citenamefont
  {Vikman}(2006)}]{Mukhanov:2005bu}%
  \BibitemOpen
  \bibfield  {author} {\bibinfo {author} {\bibfnamefont {Viatcheslav~F.}\
  \bibnamefont {Mukhanov}}\ and\ \bibinfo {author} {\bibfnamefont {Alexander}\
  \bibnamefont {Vikman}},\ }\bibfield  {title} {\enquote {\bibinfo {title}
  {Enhancing the tensor-to-scalar ratio in simple inflation},}\ }\href
  {\doibase 10.1088/1475-7516/2006/02/004} {\bibfield  {journal} {\bibinfo
  {journal} {JCAP}\ }\textbf {\bibinfo {volume} {02}},\ \bibinfo {pages} {004}
  (\bibinfo {year} {2006})},\ \Eprint {http://arxiv.org/abs/astro-ph/0512066}
  {arXiv:astro-ph/0512066} \BibitemShut {NoStop}%
\bibitem [{\citenamefont {Babichev}\ \emph {et~al.}(2006)\citenamefont
  {Babichev}, \citenamefont {Mukhanov},\ and\ \citenamefont
  {Vikman}}]{Babichev:2006vx}%
  \BibitemOpen
  \bibfield  {author} {\bibinfo {author} {\bibfnamefont {E.}~\bibnamefont
  {Babichev}}, \bibinfo {author} {\bibfnamefont {Viatcheslav~F.}\ \bibnamefont
  {Mukhanov}}, \ and\ \bibinfo {author} {\bibfnamefont {A.}~\bibnamefont
  {Vikman}},\ }\bibfield  {title} {\enquote {\bibinfo {title} {Escaping from
  the black hole?}}\ }\href {\doibase 10.1088/1126-6708/2006/09/061} {\bibfield
   {journal} {\bibinfo  {journal} {JHEP}\ }\textbf {\bibinfo {volume} {09}},\
  \bibinfo {pages} {061} (\bibinfo {year} {2006})},\ \Eprint
  {http://arxiv.org/abs/hep-th/0604075} {arXiv:hep-th/0604075} \BibitemShut
  {NoStop}%
\bibitem [{\citenamefont {Babichev}\ \emph {et~al.}(2007)\citenamefont
  {Babichev}, \citenamefont {Mukhanov},\ and\ \citenamefont
  {Vikman}}]{Babichev:2007wg}%
  \BibitemOpen
  \bibfield  {author} {\bibinfo {author} {\bibfnamefont {Eugeny}\ \bibnamefont
  {Babichev}}, \bibinfo {author} {\bibfnamefont {Viatcheslav}\ \bibnamefont
  {Mukhanov}}, \ and\ \bibinfo {author} {\bibfnamefont {Alexander}\
  \bibnamefont {Vikman}},\ }\bibfield  {title} {\enquote {\bibinfo {title}
  {Looking beyond the horizon},}\ }in\ \href {\doibase
  10.1142/9789812834300_0171} {\emph {\bibinfo {booktitle} {{11th Marcel
  Grossmann Meeting on General Relativity}}}}\ (\bibinfo {year} {2007})\ pp.\
  \bibinfo {pages} {1471--1474},\ \Eprint {http://arxiv.org/abs/0704.3301}
  {arXiv:0704.3301 [hep-th]} \BibitemShut {NoStop}%
\bibitem [{\citenamefont {Elder}\ \emph {et~al.}(2015)\citenamefont {Elder},
  \citenamefont {Joyce}, \citenamefont {Khoury},\ and\ \citenamefont
  {Tolley}}]{positivity}%
  \BibitemOpen
  \bibfield  {author} {\bibinfo {author} {\bibfnamefont {Benjamin}\
  \bibnamefont {Elder}}, \bibinfo {author} {\bibfnamefont {Austin}\
  \bibnamefont {Joyce}}, \bibinfo {author} {\bibfnamefont {Justin}\
  \bibnamefont {Khoury}}, \ and\ \bibinfo {author} {\bibfnamefont {Andrew~J.}\
  \bibnamefont {Tolley}},\ }\bibfield  {title} {\enquote {\bibinfo {title}
  {{Positive energy theorem for $P(X,\phi)$ theories}},}\ }\href {\doibase
  10.1103/PhysRevD.91.064002} {\bibfield  {journal} {\bibinfo  {journal} {Phys.
  Rev. D}\ }\textbf {\bibinfo {volume} {91}},\ \bibinfo {pages} {064002}
  (\bibinfo {year} {2015})},\ \Eprint {http://arxiv.org/abs/1405.7696}
  {arXiv:1405.7696 [hep-th]} \BibitemShut {NoStop}%
\bibitem [{\citenamefont {Greiter}\ \emph {et~al.}(1989)\citenamefont
  {Greiter}, \citenamefont {Wilczek},\ and\ \citenamefont
  {Witten}}]{Greiter:1989qb}%
  \BibitemOpen
  \bibfield  {author} {\bibinfo {author} {\bibfnamefont {Martin}\ \bibnamefont
  {Greiter}}, \bibinfo {author} {\bibfnamefont {Frank}\ \bibnamefont
  {Wilczek}}, \ and\ \bibinfo {author} {\bibfnamefont {Edward}\ \bibnamefont
  {Witten}},\ }\bibfield  {title} {\enquote {\bibinfo {title} {Hydrodynamic
  relations in superconductivity},}\ }\href {\doibase
  10.1142/S0217984989001400} {\bibfield  {journal} {\bibinfo  {journal} {Mod.
  Phys. Lett.}\ }\textbf {\bibinfo {volume} {B3}},\ \bibinfo {pages} {903}
  (\bibinfo {year} {1989})}\BibitemShut {NoStop}%
\bibitem [{\citenamefont {Son}(2001)}]{Son:2000ht}%
  \BibitemOpen
  \bibfield  {author} {\bibinfo {author} {\bibfnamefont {D.~T.}\ \bibnamefont
  {Son}},\ }\bibfield  {title} {\enquote {\bibinfo {title} {Hydrodynamics of
  relativistic systems with broken continuous symmetries},}\ }\bibfield
  {booktitle} {\emph {\bibinfo {booktitle} {{Particles and fields. Proceedings,
  Meeting, DPF 2000, Columbus, USA, August 9-12, 2000}}},\ }\href {\doibase
  10.1142/S0217751X01009545} {\bibfield  {journal} {\bibinfo  {journal} {Int.
  J. Mod. Phys.}\ }\textbf {\bibinfo {volume} {A16S1C}},\ \bibinfo {pages}
  {1284--1286} (\bibinfo {year} {2001})},\ \Eprint
  {http://arxiv.org/abs/hep-ph/0011246} {arXiv:hep-ph/0011246 [hep-ph]}
  \BibitemShut {NoStop}%
\bibitem [{\citenamefont {Son}(2002)}]{Son:2002zn}%
  \BibitemOpen
  \bibfield  {author} {\bibinfo {author} {\bibfnamefont {D.~T.}\ \bibnamefont
  {Son}},\ }\bibfield  {title} {\enquote {\bibinfo {title} {Low-energy quantum
  effective action for relativistic superfluids},}\ }\href@noop {} {\
  (\bibinfo {year} {2002})},\ \Eprint {http://arxiv.org/abs/hep-ph/0204199}
  {arXiv:hep-ph/0204199 [hep-ph]} \BibitemShut {NoStop}%
\bibitem [{\citenamefont {Ballesteros}\ and\ \citenamefont
  {Bellazzini}(2013)}]{self-gravitating-fluid}%
  \BibitemOpen
  \bibfield  {author} {\bibinfo {author} {\bibfnamefont {Guillermo}\
  \bibnamefont {Ballesteros}}\ and\ \bibinfo {author} {\bibfnamefont {Brando}\
  \bibnamefont {Bellazzini}},\ }\bibfield  {title} {\enquote {\bibinfo {title}
  {{Effective perfect fluids in cosmology}},}\ }\href {\doibase
  10.1088/1475-7516/2013/04/001} {\bibfield  {journal} {\bibinfo  {journal}
  {JCAP}\ }\textbf {\bibinfo {volume} {04}},\ \bibinfo {pages} {001} (\bibinfo
  {year} {2013})},\ \Eprint {http://arxiv.org/abs/1210.1561} {arXiv:1210.1561
  [hep-th]} \BibitemShut {NoStop}%
\bibitem [{\citenamefont {Dubovsky}\ \emph {et~al.}(2006)\citenamefont
  {Dubovsky}, \citenamefont {Gregoire}, \citenamefont {Nicolis},\ and\
  \citenamefont {Rattazzi}}]{Dubovsky:2005xd}%
  \BibitemOpen
  \bibfield  {author} {\bibinfo {author} {\bibfnamefont {S.}~\bibnamefont
  {Dubovsky}}, \bibinfo {author} {\bibfnamefont {T.}~\bibnamefont {Gregoire}},
  \bibinfo {author} {\bibfnamefont {A.}~\bibnamefont {Nicolis}}, \ and\
  \bibinfo {author} {\bibfnamefont {R.}~\bibnamefont {Rattazzi}},\ }\bibfield
  {title} {\enquote {\bibinfo {title} {Null energy condition and superluminal
  propagation},}\ }\href {\doibase 10.1088/1126-6708/2006/03/025} {\bibfield
  {journal} {\bibinfo  {journal} {JHEP}\ }\textbf {\bibinfo {volume} {03}},\
  \bibinfo {pages} {025} (\bibinfo {year} {2006})},\ \Eprint
  {http://arxiv.org/abs/hep-th/0512260} {arXiv:hep-th/0512260} \BibitemShut
  {NoStop}%
\bibitem [{\citenamefont {Andersson}\ and\ \citenamefont
  {Comer}(2020)}]{Andersson:2020uue}%
  \BibitemOpen
  \bibfield  {author} {\bibinfo {author} {\bibfnamefont {N.}~\bibnamefont
  {Andersson}}\ and\ \bibinfo {author} {\bibfnamefont {G.L.C.}\ \bibnamefont
  {Comer}},\ }\bibfield  {title} {\enquote {\bibinfo {title} {Relativistic
  fluid dynamics: physics for many different scales},}\ }\href@noop {} {\
  (\bibinfo {year} {2020})},\ \Eprint {http://arxiv.org/abs/2008.12069}
  {arXiv:2008.12069 [gr-qc]} \BibitemShut {NoStop}%
\bibitem [{\citenamefont {Endlich}\ \emph {et~al.}(2013)\citenamefont
  {Endlich}, \citenamefont {Nicolis},\ and\ \citenamefont
  {Wang}}]{self-gravitating-infl}%
  \BibitemOpen
  \bibfield  {author} {\bibinfo {author} {\bibfnamefont {Solomon}\ \bibnamefont
  {Endlich}}, \bibinfo {author} {\bibfnamefont {Alberto}\ \bibnamefont
  {Nicolis}}, \ and\ \bibinfo {author} {\bibfnamefont {Junpu}\ \bibnamefont
  {Wang}},\ }\bibfield  {title} {\enquote {\bibinfo {title} {{Solid
  Inflation}},}\ }\href {\doibase 10.1088/1475-7516/2013/10/011} {\bibfield
  {journal} {\bibinfo  {journal} {JCAP}\ }\textbf {\bibinfo {volume} {10}},\
  \bibinfo {pages} {011} (\bibinfo {year} {2013})},\ \Eprint
  {http://arxiv.org/abs/1210.0569} {arXiv:1210.0569 [hep-th]} \BibitemShut
  {NoStop}%
\bibitem [{\citenamefont {Gruzinov}(2004)}]{Gruzinov:2004ty}%
  \BibitemOpen
  \bibfield  {author} {\bibinfo {author} {\bibfnamefont {Andrei}\ \bibnamefont
  {Gruzinov}},\ }\bibfield  {title} {\enquote {\bibinfo {title} {Elastic
  inflation},}\ }\href {\doibase 10.1103/PhysRevD.70.063518} {\bibfield
  {journal} {\bibinfo  {journal} {Phys. Rev. D}\ }\textbf {\bibinfo {volume}
  {70}},\ \bibinfo {pages} {063518} (\bibinfo {year} {2004})},\ \Eprint
  {http://arxiv.org/abs/astro-ph/0404548} {arXiv:astro-ph/0404548} \BibitemShut
  {NoStop}%
\bibitem [{\citenamefont {Babichev}(2016)}]{Babichev:2016hys}%
  \BibitemOpen
  \bibfield  {author} {\bibinfo {author} {\bibfnamefont {Eugeny}\ \bibnamefont
  {Babichev}},\ }\bibfield  {title} {\enquote {\bibinfo {title} {Formation of
  caustics in k-essence and horndeski theory},}\ }\href {\doibase
  10.1007/JHEP04(2016)129} {\bibfield  {journal} {\bibinfo  {journal} {JHEP}\
  }\textbf {\bibinfo {volume} {04}},\ \bibinfo {pages} {129} (\bibinfo {year}
  {2016})},\ \Eprint {http://arxiv.org/abs/1602.00735} {arXiv:1602.00735
  [hep-th]} \BibitemShut {NoStop}%
\bibitem [{\citenamefont {Frolov}\ \emph {et~al.}(2002)\citenamefont {Frolov},
  \citenamefont {Kofman},\ and\ \citenamefont {Starobinsky}}]{Frolov:2002rr}%
  \BibitemOpen
  \bibfield  {author} {\bibinfo {author} {\bibfnamefont {Andrei~V.}\
  \bibnamefont {Frolov}}, \bibinfo {author} {\bibfnamefont {Lev}\ \bibnamefont
  {Kofman}}, \ and\ \bibinfo {author} {\bibfnamefont {Alexei~A.}\ \bibnamefont
  {Starobinsky}},\ }\bibfield  {title} {\enquote {\bibinfo {title} {Prospects
  and problems of tachyon matter cosmology},}\ }\href {\doibase
  10.1016/S0370-2693(02)02582-0} {\bibfield  {journal} {\bibinfo  {journal}
  {Phys. Lett.}\ }\textbf {\bibinfo {volume} {B545}},\ \bibinfo {pages} {8--16}
  (\bibinfo {year} {2002})},\ \Eprint {http://arxiv.org/abs/hep-th/0204187}
  {arXiv:hep-th/0204187 [hep-th]} \BibitemShut {NoStop}%
\bibitem [{\citenamefont {Bilic}(2008{\natexlab{a}})}]{Bilic:2008zz}%
  \BibitemOpen
  \bibfield  {author} {\bibinfo {author} {\bibfnamefont {Neven}\ \bibnamefont
  {Bilic}},\ }\bibfield  {title} {\enquote {\bibinfo {title} {Thermodynamics of
  dark energy},}\ }\href {\doibase 10.1002/prop.200710507} {\bibfield
  {journal} {\bibinfo  {journal} {Fortsch. Phys.}\ }\textbf {\bibinfo {volume}
  {56}},\ \bibinfo {pages} {363--372} (\bibinfo {year} {2008}{\natexlab{a}})},\
  \Eprint {http://arxiv.org/abs/0812.5050} {arXiv:0812.5050 [gr-qc]}
  \BibitemShut {NoStop}%
\bibitem [{\citenamefont {Bilic}(2008{\natexlab{b}})}]{Bilic:2008zk}%
  \BibitemOpen
  \bibfield  {author} {\bibinfo {author} {\bibfnamefont {Neven}\ \bibnamefont
  {Bilic}},\ }\bibfield  {title} {\enquote {\bibinfo {title} {Thermodynamics of
  k-essence},}\ }\href {\doibase 10.1103/PhysRevD.78.105012} {\bibfield
  {journal} {\bibinfo  {journal} {Phys. Rev. D}\ }\textbf {\bibinfo {volume}
  {78}},\ \bibinfo {pages} {105012} (\bibinfo {year} {2008}{\natexlab{b}})},\
  \Eprint {http://arxiv.org/abs/0806.0642} {arXiv:0806.0642 [gr-qc]}
  \BibitemShut {NoStop}%
\bibitem [{\citenamefont {Tolley}\ and\ \citenamefont
  {Wyman}(2010)}]{Tolley:2009fg}%
  \BibitemOpen
  \bibfield  {author} {\bibinfo {author} {\bibfnamefont {Andrew~J.}\
  \bibnamefont {Tolley}}\ and\ \bibinfo {author} {\bibfnamefont {Mark}\
  \bibnamefont {Wyman}},\ }\bibfield  {title} {\enquote {\bibinfo {title} {The
  gelaton scenario: Equilateral non-gaussianity from multi-field dynamics},}\
  }\href {\doibase 10.1103/PhysRevD.81.043502} {\bibfield  {journal} {\bibinfo
  {journal} {Phys. Rev. D}\ }\textbf {\bibinfo {volume} {81}},\ \bibinfo
  {pages} {043502} (\bibinfo {year} {2010})},\ \Eprint
  {http://arxiv.org/abs/0910.1853} {arXiv:0910.1853 [hep-th]} \BibitemShut
  {NoStop}%
\bibitem [{\citenamefont {Babichev}\ and\ \citenamefont
  {Ramazanov}(2017)}]{Babichev:2017lrx}%
  \BibitemOpen
  \bibfield  {author} {\bibinfo {author} {\bibfnamefont {Eugeny}\ \bibnamefont
  {Babichev}}\ and\ \bibinfo {author} {\bibfnamefont {Sabir}\ \bibnamefont
  {Ramazanov}},\ }\bibfield  {title} {\enquote {\bibinfo {title} {Caustic free
  completion of pressureless perfect fluid and k-essence},}\ }\href {\doibase
  10.1007/JHEP08(2017)040} {\bibfield  {journal} {\bibinfo  {journal} {JHEP}\
  }\textbf {\bibinfo {volume} {08}},\ \bibinfo {pages} {040} (\bibinfo {year}
  {2017})},\ \Eprint {http://arxiv.org/abs/1704.03367} {arXiv:1704.03367
  [hep-th]} \BibitemShut {NoStop}%
\bibitem [{\citenamefont {Babichev}\ \emph {et~al.}(2018)\citenamefont
  {Babichev}, \citenamefont {Ramazanov},\ and\ \citenamefont
  {Vikman}}]{Babichev:2018twg}%
  \BibitemOpen
  \bibfield  {author} {\bibinfo {author} {\bibfnamefont {Eugeny}\ \bibnamefont
  {Babichev}}, \bibinfo {author} {\bibfnamefont {Sabir}\ \bibnamefont
  {Ramazanov}}, \ and\ \bibinfo {author} {\bibfnamefont {Alexander}\
  \bibnamefont {Vikman}},\ }\bibfield  {title} {\enquote {\bibinfo {title}
  {{Recovering $P(X)$ from a canonical complex field}},}\ }\href {\doibase
  10.1088/1475-7516/2018/11/023} {\bibfield  {journal} {\bibinfo  {journal}
  {JCAP}\ }\textbf {\bibinfo {volume} {11}},\ \bibinfo {pages} {023} (\bibinfo
  {year} {2018})},\ \Eprint {http://arxiv.org/abs/1807.10281} {arXiv:1807.10281
  [gr-qc]} \BibitemShut {NoStop}%
\bibitem [{\citenamefont {Mukohyama}\ and\ \citenamefont
  {Namba}(2020)}]{Mukohyama:2020lsu}%
  \BibitemOpen
  \bibfield  {author} {\bibinfo {author} {\bibfnamefont {Shinji}\ \bibnamefont
  {Mukohyama}}\ and\ \bibinfo {author} {\bibfnamefont {Ryo}\ \bibnamefont
  {Namba}},\ }\bibfield  {title} {\enquote {\bibinfo {title} {{Partial UV
  Completion of $P(X)$ from a Curved Field Space}},}\ }\href@noop {} {\
  (\bibinfo {year} {2020})},\ \Eprint {http://arxiv.org/abs/2010.09184}
  {arXiv:2010.09184 [hep-th]} \BibitemShut {NoStop}%
\end{thebibliography}%

\end{document}